\documentclass{aa}

\newcommand{\unit}[1]{\; \mbox{#1}}
\newcommand{\un}[1]{_{\mbox{\scriptsize #1}}}

\usepackage{epsfig}
\usepackage{amssymb}
\usepackage{graphicx}

\begin{document}

\onecolumn

\title{$\beta$--Radioactive Cosmic Rays in a diffusion model: test for a
local
bubble?}
\author{Fiorenza Donato
       \inst{1}
       \and David Maurin
       \inst{1,2}
       \and Richard Taillet
       \inst{1,2}
       }

       \authorrunning{Donato, Maurin \& Taillet}
       \titlerunning{Radioactive Cosmic Rays Diffusive Propagation}
\institute{
Laboratoire de Physique Th\'eorique {\sc lapth},
Annecy--le--Vieux, 74941, France
\and
Universit\'e de Savoie, Chamb\'ery, 73011, France
}

\date{Received 7 August 2001; Accepted 12 September 2001}

\abstract{
In a previous analysis,
Maurin et al. (2001) have constrained
several parameters  of the cosmic ray  diffusive
propagation (the diffusion
coefficient normalization $K_0$ and its spectral index $\delta$, the
halo
half-thickness $L$, the Alfv\'en velocity $V_a$,
and the galactic wind $V_c$)
using stable nuclei.
In a second paper (Donato et al., 2001),
these parameters were
shown to reproduce the observed antiproton spectrum with no further
adjustment.
In the present paper, we extend the analysis to the $\beta$-radioactive
nuclei $^{10}$Be, $^{26}$Al and $^{36}$Cl.
These species will be shown to be particularly sensitive to the
properties of the local interstellar medium ({\sc lism}).
As studies of the {\sc lism} suggest that we live in an underdense
bubble
of extent $r_{hole} \sim 50 - 200 \unit{pc}$,
this local feature must be taken into account.
We present a modified version of our diffusion model which
describes the underdensity as a hole in the galactic disc;
we believe some of the formul\ae \ presented here are new.
It is found that the presence of the bubble leads to a decrease in
the radioactive fluxes which can be approximated by a simple factor
$\exp(-r_{hole}/l_{rad})$ where $l_{rad}=\sqrt{K \gamma \tau_0}$ is the
typical distance travelled by a radioactive nucleus before it decays.
We find that each of the radioactive nuclei independently points
towards a bubble of radius $\lesssim 100 \unit{pc}$.
If these nuclei are considered simultaneously, only models with a bubble
radius $r_{hole} \sim 60 - 100 \unit{pc}$ are marginally consistent with
data. In particular, the standard case $r_{hole}=0 \unit{pc}$ is
disfavored.
Our main concern is about the consistency
of the currently available data, especially $^{26}$Al/$^{27}$Al.
\keywords{Cosmic rays -- Local interstellar medium}
}

\maketitle

\section{Introduction}

The idea of using long-lived radioactive isotopes to estimate the
average age of
cosmic rays is not new and the name "radioactive chronometer" is now
standard. Indeed, the finite lifetime of these nuclei has a deep
impact on their propagation and they are expected to give independent
clues about the propagation parameters.
To this end, the nucleus $^{10}$Be ($t_{1/2}=1.51$ Myr) was proposed
in the fifties, but due to an insufficient knowledge of the processes
occurring during
propagation, it has long been  thought not to be the best candidate
after all.
Instead, Reames (1980)
proposed $^{53}$Mn which is {\sc ec} (electronic capture)
unstable ($t_{1/2}=3.74$ Myr), but
it was subsequently shown to be of no great interest, owing to
the specific behavior of the {\sc ec} channel. Later on,
Cass\'e (1973)
proposed $^{54}$Mn
which is a mixed $\beta$-{\sc ec} unstable as a chronometer of the Fe
group,
the latter being mostly created by collisions of cosmic ray Fe on
the interstellar medium. These propositions are very instructive since they
demonstrate the early hesitations and progressive comprehension of
cosmic ray nuclei propagation. From an observational
point of view, the first clues about this propagation time came
from the isotope $^{10}$Be  (Hagen et al., 1977; Webber et al., 1977;
Garcia-Munoz et al., 1977). Other isotopically resolved chronometers
followed as $^{26}$Al (Freier et al., 1980), $^{36}$Cl (Wiedenbeck,
1985)
and $^{54}$Mn (Leske, 1993). Since then, numerous accurate experiments
have been performed (see section~\ref{sec:dataset}) giving more confident
estimates of the surviving fraction of these radioactive clocks.

Most of the conclusions of radioactive studies have been drawn in
leaky box models, using the notion of surviving fraction to estimate
the age of cosmic rays.
The early analyses focused on the surviving fraction and this
quantity hinted at the existence of a diffusive halo far
beyond the thin galactic disc (see in particular Simpson and
Garcia-Munoz, 1987 for a discussion).
We emphasize that in diffusion models, these notions have no clear
physical meaning (see discussion in section~\ref{Global-local}), as
stable species and each radioactive species have different ages.
The main goal nowadays is rather to use a diffusion model that
can consistently reproduce all observed radioactive abundances.
Our purpose here -- following Ptuskin et al. (1997), Ptuskin and
Soutoul (1998) -- is to demonstrate that one may have
to consider local properties of the interstellar medium ({\sc ism})
in order to bring this program to a successful conclusion.

The plan of the paper is the following: in a first part we
recall the nuclear properties of the radioactive species.
In the second part, we focus on the astrophysical
aspects of propagation.
More specifically, we introduce an inhomogeneous
model for the production and
propagation of radioactive species.
We also discuss the meaning of older models and some false claims that
are made in the literature about the predictive power of $\beta$
radioactive clocks.
We then present our analysis and results.


\section{Description of radioactive species}

From the propagation point of view, one can associate three attributes
to each nucleus: (i) a reaction cross section,
(ii) a spallation production cross section and
(iii) a rest frame half-life.
We will not discuss further the first two points
as they were discussed at length in Maurin et al. (2001) (hereafter
Paper I).
Indeed, the cross section code and parameterizations
developed in Webber et al. (1990),
Webber et al. (1998) (spallation cross section)
and Tripathi et al. (1997a; 1997b; 1999) (reaction cross section) are
appropriate for both stable and radioactive nuclei.

However, it is not so straightforward to specify decay channels
for unstable nuclei. Despite the fact that one can find half-lives
easily in nuclear charts (Audi et al., 1997)\footnote{see also
http://sutekh.nd.rl.ac.uk/CoN/ or
http://nucleardata.nuclear.lu.se/nucleardata/}, the information needed
from a cosmicist point of view is not always clearly extractable: for
example $\beta$ and {\sc ec} modes are often combined whereas these
two effects act differently during propagation.
Thus in the following section, we give a list of nuclei that need to be
propagated as unstable. We emphasize that these
ingredients were already present in the version of our code
used in Paper I, although they were not explicitly discussed.

\subsection{Half-lives}
Two different disintegration modes must be distinguished:
$\beta$ decay (the nucleus decays spontaneously with a given
lifetime $\tau_0 = t_{1/2} / \ln 2$)
and {\sc ec} decay (the nucleus decays after capturing a K-shell
electron if one has been attached).
In the latter case, the decay rate depends on the
electron density and on the typical attachment time.
On Earth, many electrons
are available so that a nucleus unstable {\em via} electronic capture
behaves as $\beta^+$ (at least in terms of the daughter nucleus and
half--life). Nevertheless
in the {\sc ism}, where fewer electrons can be attached, the attachment
time must be taken into account in the {\sc ec} mode.
We will come back to this point later on.

Among all the radioactive nuclei, those which have
lifetimes of the order of the propagation time may bring interesting
information about their propagation.
We select and classify them into three categories:
$\beta$--unstable, {\sc ec}--unstable and mixed $\beta$--{\sc
ec}--unstable.
We emphasize that this classification only has a meaning
in the context of cosmic ray propagation: were we interested
in stellar nucleosynthesis, for example, other tables would be necessary
to encode disintegration channels as the characteristic times
of evolution involved and the properties of the medium would be
drastically
different.

Up to now, the most complete description
of half-lives adapted to cosmic ray description has been given in
Letaw et al. (1984).
In this paper, we only consider $\beta$
unstable nuclei, but as results about {\sc ec} species
will be presented in a separate paper (Donato et al. 2001, in
preparation)
we find it convenient to describe in this section all unstable species
($\beta$ and {\sc ec}). Thus, the following tables are
similar to those in Letaw et al. (1984), but we restrict ourselves
to $Z\leq 30$ and we take the opportunity to incorporate
updated values for some half-lives.

\subsubsection{\boldmath{$\beta$} unstable}
\label{sec:beta_pur}
In Table \ref{TableAU BETA-}, we present the three pure $\beta$ unstable
isotopes having proper half-lives in the time range 1~kyr-100~Myr.
Nucleus $^{14}$C
is the shortest-living $\beta$ unstable isotope included in our cascade.
This choice is motivated by the measured propagation time, which runs
between 10-20~Myr depending on the propagation model (this time
decreases with energy). If a nucleus has a half-life $\lesssim$ kyr,
we can consider it as a ghost nucleus (see details in Paper I); 
this means in particular that as soon as it is created,
this nucleus immediately (compared to the typical propagation time)
disappears into its daughter nucleus, so that its density
in cosmic radiation is negligible and it does not need to be propagated.
\begin{table}[hbt!]
    \caption{Pure $\beta$ unstable isotopes
    ($1\unit{kyr}<t_{1/2}<100\unit{Myr}$).}
    \label{TableAU BETA-}
    \begin{center}
	\begin{tabular}{|c|c|c|c|c|}   \hline
	    $Z$  & Nucleus  &  Daughter    & $t^{unit.}_{1/2}$(error)\\
	    \hline \hline
	    \vbox{\vspace{0.4cm}}
	    $4$  &  $^{10}_{4}$Be  & $^{10}_{5}$B & $1.51^{Myr}\;(0.06)$ \\
	    \vbox{\vspace{0.4cm}}
	    $6$ & $^{14}_{6}$C  & $^{14}_{7}$N & $5.73^{kyr}\;(0.04)$\\
	    \vbox{\vspace{0.4cm}}
	    $26$ & $^{60}_{26}$Fe$^{\dagger}$ & $(^{60}_{27}$Co$
	    \stackrel{\beta^-}{\rightarrow})^{60}_{28}$Ni & $1.5^{Myr}\;(0.3)$ \\ 
	    \hline \hline
	\end{tabular}
    \end{center}
    {\footnotesize
    $^{\dagger}$ The transition from Fe to Co has a half-life of
    $1.5$ Myr while transition from Co to Ni is immediate
    from a cosmic ray point of view ($t_{1/2}\sim 5$ yr).}
\end{table}
We might object our $\tau_0 >1
\unit{kpc}$ cut-off, arguing that because of the Lorentz factor
$\gamma = E_{tot}/m$, even a short
proper lifetime  can give a large effective lifetime at
sufficiently high energy.
Actually, we are concerned with energies $E_{tot} < 100
\unit{GeV/nuc}$, so that $\gamma < 100$. It follows that
proper half-lives $\tau_0 < 1 \unit{kyr}$ correspond to effective
half-lives $\tau = \gamma \tau_0 < 0.1 \unit{Myr}$, which is too short
compared to the propagation time for our purpose.
Moreover, this cut-off is natural in practice as
there are no radioactive half-lives in the interval 0.1-1~kyr.
Conversely, if this lifetime is too high, the corresponding nucleus
could be
propagated as a stable one. To give an example, $^{40}$K
-- whose $\beta$ half-life is about $\sim 10^9$ yr --
can clearly be considered as stable.
\subsubsection{EC unstable}
In Table \ref{TableAU EC}, we  present the isotopes that
are unstable under electronic capture transitions.
Unlike  for $\beta$-decay, these nuclei have to attach an electron to
be able to decay.
As (i) nuclei are completely stripped of $e^-$  at cosmic ray energies
and (ii) the interstellar medium is very poor in $e^-$,
the attachment time may be much longer than the lifetime of the
decay. This means that even a nucleus with a very short
{\sc ec}--lifetime may have a long effective lifetime.
As a result, there is no need to consider any lower bound
on the half-lives.
This will be discussed further in
section~\ref{sec:Decay-prop}, and
the interested reader can refer to Letaw et al (1984) or
Adams et al (1985) who show in a leaky box model that the effective
half-life
of {\sc ec} nuclei is at least of the order of the attachment rate,
which is
about $\sim$ Myr$^{-1}$.

\begin{table}[hbt!]
    \caption{Pure {\sc ec} unstable isotopes.}
    \label{TableAU EC}
    \begin{center}
	\begin{tabular}{|c|c|c|c|c|}   \hline
	    $Z$  &  unstable ({\sc ec})  &  Daughter &
	    $t^{unit.}_{1/2}$
	    (error)\\ \hline \hline
	    \vbox{\vspace{0.4cm}}
	    $4$ &  $^{7}_{4}$Be & $^{7}_{3}$Li & $ 53.29^d\;(0.07)$ \\
	    \vbox{\vspace{0.5cm}}
	    $18$ & $^{37}_{18}$Ar & $^{37}_{17}$Cl  & $35.04^d
	    \;(0.04)$\\
	    \vbox{\vspace{0.5cm}}
	    $20$ & $^{41}_{20}$Ca & $^{41}_{19}$K  & $103^{kyr}\;(4)$\\
	    \vbox{\vspace{0.5cm}}
	    $22$ & $^{44}_{22}$Ti& $(^{44}_{21}$Sc$
	    \stackrel{\beta^+}{\rightarrow})
	    ^{44}_{20}$Ca  & $49^{yr}\;(3)$\\
	    \vbox{\vspace{0.5cm}}
	    $23$ & $^{49}_{23}$V & $^{49}_{22}$Ti & $330^d\; (15)$ \\
	    \vbox{\vspace{0.5cm}}
	    $24$ & $^{48}_{24}$Cr$^{\dagger}$ & $(^{48}_{23}$V$
	    \stackrel{\beta^+}{\rightarrow})
	    ^{48}_{22}$Ti &
	    $21.56^h\;(0.03)^{\ddagger}$\\
	    \vbox{\vspace{0.5cm}}
	    $24$ & $^{51}_{24}$Cr & $^{51}_{23}$V & $27.702^d
	    \;(0.004)$ \\
	    \vbox{\vspace{0.5cm}}
	    $25$ & $^{53}_{25}$Mn &$^{53}_{24}$Cr &
	    $3.74^{Myr}\;(0.04)$ \\
	    \vbox{\vspace{0.5cm}}
	    $26$ & $^{55}_{26}$Fe & $^{55}_{25}$Mn & $2.73^{yr}
	    \;(0.03)$ \\
	    \vbox{\vspace{0.5cm}}
	    $27$ & $^{57}_{27}$Co & $^{57}_{26}$Fe & $271.79^d
	    \;(0.09)$ \\
	    \vbox{\vspace{0.5cm}}
	    $28$ & $^{59}_{28}$Ni$^{\S}$ & $^{59}_{27}$Co    &
	    $80^{kyr}\;(11)$ \\ \hline \hline
	\end{tabular}
    \end{center}
    {\footnotesize
    $^{\dagger}$ This nucleus has an allowed $\beta$ transition,
    but contrary to $^{54}$Mn and $^{56}$Ni 
    (cf Table~\ref{DOUBLE MODE DE DESINTEGRATION}),
    it has not been studied recently so that we can set it as a pure {\sc ec}.\\
    $^{\ddagger}$ In this two-step reaction, the second transition
    $^{48}$V$\stackrel{{\beta^+}}{\rightarrow}$$^{48}$Ti has
    a half-life greater than the first one -- $15.9735^d(0.0025)$.
    Nevertheless, this second reaction can be taken as immediate because 
    of its $\beta$ nature.
    We thus can consider this second element as a ghost (see text).
    Finally, only the first reaction ($^{48}$Cr$\rightarrow ^{48}$V)
    enters the decay rate.\\
    $^{\S}$ This nucleus has a $\beta$ decay, but  with
    $t_{1/2}>100\unit{Gyr}$. For the same reason as explained in
    section~\ref{sec:beta_pur},
    it is sufficient to take into account the {\sc ec} channel.}
\end{table}

\subsubsection{Mixed $\beta$--EC unstable}
Finally we turn to the case in which the two decay modes
are allowed (hereafter mixed decay), in table~\ref{DOUBLE MODE DE DESINTEGRATION}.
In general, $\beta$ decay is dominant,
but we also need to consider some {\sc ec} contributions.
\begin{table}[hbt!]
    \caption{Mixed {\sc ec}-$\beta$ isotopes}
    \label{DOUBLE MODE DE DESINTEGRATION}
    \begin{center}
	\begin{tabular}{|c|c||c|c||c|c|}   \hline
	    $Z$  & Nucleus  &  Daughter ({\sc ec}) & $t^{unit.}_{1/2}$
	    (error) &  Daughter
	    ($\beta$) & $t^{unit.}_{1/2}$ (error)\\ \hline \hline
	    \vbox{\vspace{0.4cm}}
	    $13$  &  $^{26}_{13}$Al$^*$  & $^{26}_{12}$Mg &
	    $4.08^{Myr}\;(0.15)$
	    & $^{26}_{12}$Mg &
	    $0.91^{Myr}\;(0.04)$\\
	    \vbox{\vspace{0.4cm}}
	    $17$ & $^{36}_{17}$Cl  & $^{36}_{16}$S &
	    $15.84^{Myr}\;(0.11)$ & $^{36}_{18}$Ar &
	    $0.307^{Myr}\;(0.002)$ \\
	    \vbox{\vspace{0.4cm}} $25$  &  $^{54}_{25}$Mn$^\dagger$  
	    & $^{54}_{24}$Cr & $312.3^d\;(0.4)$ & $^{54}_{26}$Fe &
	    $0.494^{Myr}\;(0.006)$\\
	    \vbox{\vspace{0.4cm}}
	    $28$  &  $^{56}_{28}$Ni$^\ddagger$  &
	    $^{\S} \! (^{56}_{27}$Co$
	    \stackrel{\beta^+}{\rightarrow})^{56}_{26}$Fe & $6.075^d\;(0.020)$
	    & $^{\S} \! (^{56}_{27}$Co$
	    \stackrel{\beta^+}{\rightarrow})^{56}_{26}$Fe &
	    $0.051^{Myr}\;(0.022)$\\\hline\hline
	\end{tabular}
    \end{center}
    {\footnotesize
    $^*$ Mart\'{\i}nez-Pinedo and Vogel (1998) (for this
    nucleus, one can often find in the literature a half-life
    of $0.74$ Myr: it corresponds to the combined  $\beta$--{\sc ec}
    channels, and this is incorrect since these two processes
    are not equivalent during propagation).\\
    $^\dagger$ $\beta$ half-life taken from Mart\'{\i}nez-Pinedo and Vogel (1998). 
    A higher value for this channel is not excluded.\\
    $^\ddagger$ Lund Fisker et al. (1999).\\
    $^{\S}$ $^{56}_{27}$Co  decays {\em via} {\sc ec} (80\%)
    and $\beta^+$ (20\%), but as the half-life is of the order of
    two months, one can consider that the only effective
    channel is $\beta$--decay so that this nucleus vanishes
    immediately. Notice that these values are taken from
    Goldman (1982). More recent references (Audi et al., 1997)
    or nuclear charts on the web (see footnote~1) are ignored
    because they give either pure $\beta$ channel or
    pure {\sc ec} channel.}
\end{table}
This is the case for $^{54}$Mn which is often discussed as a possible chronometer for
the Fe group, but its {\sc ec} contribution is rarely treated correctly.
Incidentally, table~\ref{DOUBLE MODE DE DESINTEGRATION}
also shows that the usual two chronometers $^{26}$Al
and $^{36}$Cl have mixed decay channels. But contrary to $^{54}$Mn we
will see in the next section that the {\sc ec} channel is completely
suppressed, and it is usually neglected.

\subsection{Modifications of decay properties during propagation:
effective half-lives}
\label{sec:Decay-prop}
While $\beta$ decay is independent of the nucleus environment,
electronic capture decay depends sharply on the {\sc ism} electronic density  as
well as on the attachment and stripping cross sections.
It also follows that the energy dependence of lifetimes is more complex
in the {\sc ec} mode than in the $\beta$ mode.

This is of great importance
for the stability of nuclei, and it gives a flavor of how different
the propagation is for these two modes,
even for similar half-lives ({\em i.e.} $t_{EC}=t_{\beta}$).
In order to clarify these specific behaviors, the pure $\beta$--decay
will serve as a reference process to which {\sc ec} processes can be
compared.
The expression for the density of a $\beta$-unstable nucleus
can be found in Webber, Lee and Gupta (1992), or formul{\ae} (A8), (A9)
and (A10) of Paper I
(the disintegration rate appears explicitly in $S_i$).

As regards the other radioactive species, they actually may
be treated like $\beta$ radioactive species as it is possible
to transform their solution into that of pure $\beta$ with an effective
rate instead of the pure {\sc ec} rate or of the usual combined
$\beta$--{\sc ec} rate. An effective lifetime can be introduced,
obtained by combining $t_{EC}$, attachment and
stripping cross sections (and $t_{\beta}$ for mixed decay).
A discussion of the validity of such approximations can be found in Letaw et
al. (1984), Adams et al. (1985) and references therein.
\begin{table}[hbt!]
    \caption{Pure $\beta$ unstable isotopes
    ($1\unit{kyr}<t_{1/2}<100\unit{Myr}$)
    from a propagation point of view.}
    \label{TableAU FINAL BETA-}
    \begin{center}
	\begin{tabular}{|c|c|c|c|c|}   \hline
	    $Z$  & Nucleus  &  Daughter    & $t^{unit.}_{1/2}$(error)\\
	    \hline \hline
	    \vbox{\vspace{0.4cm}}
	    $4$  &  $^{10}_{4}$Be  & $^{10}_{5}$B &
	    $1.51^{Myr}\;(0.06)$ \\
	    \vbox{\vspace{0.4cm}}
	    $6$ & $^{14}_{6}$C  & $^{14}_{7}$N & $5.73^{kyr}\;(0.04)$\\
	    \vbox{\vspace{0.4cm}}
	    $13$  &  $^{26}_{13}$Al & $^{26}_{12}$Mg &
	    $0.91^{Myr}\;(0.04)$\\
	    \vbox{\vspace{0.4cm}}
	    $17$ & $^{36}_{17}$Cl  & $^{36}_{18}$Ar &
	    $0.307^{Myr}\;(0.002)$ \\
	    \vbox{\vspace{0.4cm}}
	    $26$ & $^{60}_{26}$Fe & $^{60}_{28}$Ni & $1.5^{Myr}\;(0.3)$\\
	    \hline \hline
	\end{tabular}
    \end{center}
\end{table}
The physical inputs required to describe {\sc
ec}--unstable and $\beta$--unstable are different and we made the
choice here to focus on the latter case. The case of {\sc ec}--unstable
species will be discussed elsewhere (Donato et al. 2001, in
preparation).
Consequently, the nuclei that one might consider as pure $\beta$ are
given in
Table~\ref{TableAU FINAL BETA-}.

We have checked that the {\sc ec} mode can be neglected for
$^{27}$Al and $^{36}$Cl but not for $^{54}$Mn and $^{56}$Ni.
The comparison of our tables to Letaw et al. (1984) shows that the
largest difference is for $^{60}$Fe half-life
which changed from $3\times10^5$ to $1.5\times10^6$ yr.
For other nuclei, minor corrections in half-lives and their
uncertainties have been made.

Finally, we recall that the propagation of nuclei which are
radioactively produced requires a specific treatment.
Their solution corresponds to formula (A11) of Paper I and as they do
not play any role in this study, we will not discuss these daughter
nuclei further.


\section{Motivation for a local treatment of radioactive cosmic rays
propagation}

The propagation of cosmic rays is  {\em a priori} affected
by the details of the gas density distribution in the galactic disc.
Several models have been proposed to take this into account:
for example, Strong and Moskalenko (1998) consider a gas density
distribution
$n(r,z)$, whereas Osborne and Ptuskin (1987)
and Ptuskin et Soutoul (1990) model the {\sc ism} as a cloudy medium.
As far as the stable species are concerned, one can use an
equivalent treatment with an average description of the {\sc ism}
density.
This is what we did in Paper I.

The situation is drastically different for radioactive
species. First, due to their finite lifetime, those that reach the solar
system must have been created locally, {\em i.e.} in a region which is
at most a few hundred parsecs away (see below).
As a consequence,  the  local interstellar matter ({\sc lism}) has to
be carefully described.

Actually, there are reasons to believe that this {\sc lism} is highly
inhomogeneous,
which motivates a more elaborate model for radioactive species.
This model will be described in the next section, after we
have discussed the locality of radioactive production and the
inhomogeneity of the {\sc lism}.

\subsection{Locality from diffusion equation}
\label{Global-local}

It is often claimed that the size of the diffusion halo can
be estimated from radioactive cosmic ray species.
This is true in the framework of a leaky box propagation model, but not
for diffusion models (see below).
Indeed, the leaky box models have been so widely used in the past that
some of their conclusions have become popular wisdom and are sometimes
used out of the proper context.
Actually, as was explicitly shown in the early seventies
(Prishchep and Ptuskin, 1975; Ginzburg et al., 1980),
leaky box models are almost never equivalent
to diffusion models, so that one should be very careful before directly
applying leaky box inspired results to another class of models, {\em
e.g.} diffusion models.

\subsubsection{The old leaky box paradigm, and what we should forget to
go further\dots}
\label{LB paradigm}

The so-called leaky box modelling of cosmic ray propagation assumes
that the particles freely move in a homogeneous finite-sized box.
When a particle reaches a boundary of the box, there is a finite
probability
that it escapes the system. As a result, forgetting about anything but
propagation, the density is given by, in this model,
\begin{displaymath}
        \frac{\partial N(E,t)}{\partial t} - \frac{N(E,t)}{\tau_{\rm
        esc}(E)} = 0
\end{displaymath}
where $\tau_{\rm  esc}(E)$ is the typical time a particle of energy
$E$ spends in the box.
It is possible to account for the measured abundance of all the
stable cosmic ray
nuclei abundances with a suitable $\tau_{\rm  esc}(E)$, adjusted for
all nuclei.
However, the physical picture provided by this model is wrong.
Cosmic rays do not freely stream in a homogeneous box, but they are
scattered by the inhomogeneities of the galactic magnetic field.
This results in a diffusive propagation.
This difference is of no importance when one is concerned only with
the local abundance of stable nuclei, and it can be shown that in
this case, there is an equivalence between these two approaches
(see discussion in Paper I and references therein).
But as soon as one tries to change the mathematically correct
local description of stable nuclei provided by the leaky box into a
physical picture and tries to extend it either to radioactive species or
to another location, trouble begins.

First, it is obvious that from the start, the leaky box model is
unable to describe the spatial distribution of cosmic rays.
We will not discuss this point further.

Second, it is also clear that this model is bound to fail for
radioactive cosmic ray nuclei.
The species we consider have a
proper lifetime of the order of $\tau_{\rm rad} \sim 1 \unit{Myr}$,
so that at the speed of
light they can travel more than $\sim 300 \unit{kpc}$ before they decay.
As the radius of the galaxy is $\sim 20 \unit{kpc}$,
these nuclei should have enough time to propagate in the whole available
volume
and their distribution is sensitive to the global geometry and
localization of the sources.
This is at variance with the diffusion model in which the average
distance travelled by a nucleus during a time $\tau_{\rm rad}$ is given
by
$\sqrt{K(E)\tau_{\rm rad}}$, which is of the order of a few hundred
parsecs at low energy. This means that in this case, the distribution
of these species only depends on the {\em local} characteristics of the
diffusive medium and of the sources.

It is sometimes claimed that a radioactive species gives
the size $L$ of the diffusive halo: the bigger
the halo, the smaller the quantity of nuclei that survive from the
sources to the Earth.
This is easily understood in the framework of the leaky box, but it is
physically wrong for diffusion models.
A more quantitative version of the intuitive argument given
above can be found in Prishchep \& Ptuskin (1975).
They showed that diffusion
models are equivalent to leaky box models only when the relationship
$\tau_0 \gg L^2/K$ is satisfied.
In the most favorable of the cases we consider here, this corresponds to
$\tau_0 \gg 100 \unit{Myr}$, which is wrong for the nuclei we
are interested in.

Thus, one should be careful not to use leaky box induced conclusions out
of this specific context. They may be very intuitive
but wrong when applied to a diffusion model.
For example, the next section shows that the radioactive-to-parent
ratio is not sensitive to the size of the halo in a diffusion model.

\subsubsection{Does a radioactive really tell something about the
halo size $L$ ?}
\label{halo_size}
In the diffusion model, a radioactive nucleus with a
proper lifetime
$\tau_0$ has most probably travelled a distance
$l_{rad} = \sqrt{K(E) \gamma \tau_0}$ between its
creation in the disc and its detection on Earth.
At low energy, this distance is much smaller than the size of the
diffusive volume, and this nucleus is not very sensitive to the
boundaries
of this volume.
This can be seen in the expression giving the ratio
radioactive/parent nucleus. In the case when there is only one parent
nucleus and if  energetic gains and losses are discarded,
the density is given by (see equation A8 of Paper I)
\begin{equation}
      N^{uns}(z=0,r)=\sum_{i=0}^{\infty}\frac{2h{\Gamma}^{prod}
      N_i^{parent}(z=0)}{A_i}J_0\left(\zeta_i\frac{r}{R}\right)
      \label{densite_rad}
\end{equation}
with
\begin{equation}
      A_i\equiv 2h {\Gamma}^{inel} +V_c +KS_i\coth (S_iL/2)
\end{equation}
and
\begin{equation}
      S_i = \left( \frac{V_c^2}{K^2} + \frac{4\zeta_i^2}{R^2}
      + \frac{4 \Gamma^{\beta}}{K}\right)^{1/2}
\end{equation}

\begin{figure}[hbt!]
\centerline{\includegraphics*[width=0.63\textwidth]{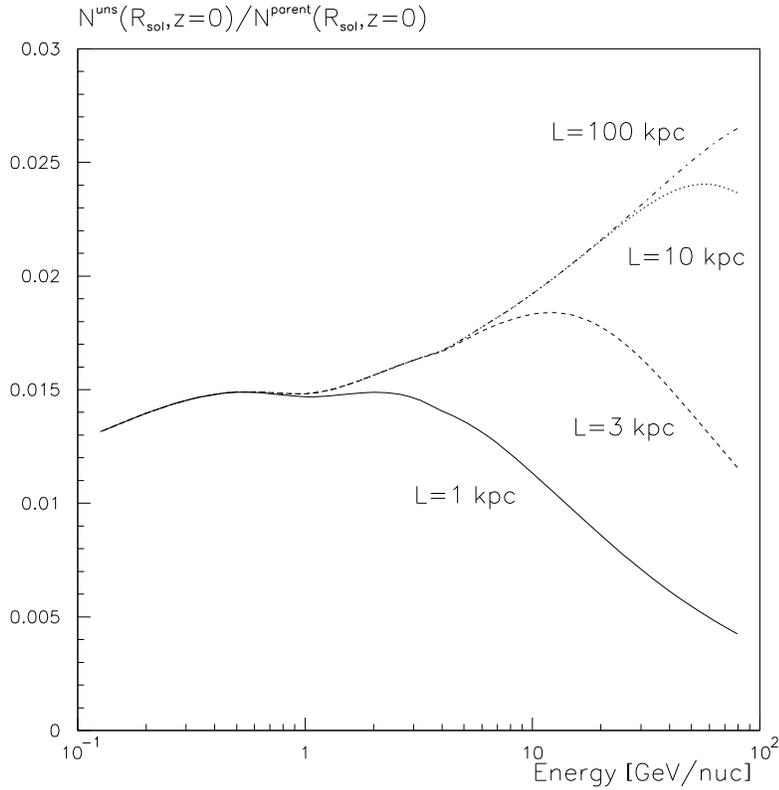}}
\caption{unstable/parent ratio ($^{26}$Al/$^{28}$Si) for different
values
of $L$ in kpc, all other parameters being fixed.}
\label{fig:influence_l}
\end{figure}

When the lifetime is so short that
$\Gamma^{\beta}L/K \approx L/l_{rad}\gg 1$, we have $\coth
(S_iL/2)\approx 1$.
It can be easily seen that in these
conditions, the ratio $N^{uns}(z=0,r)/N^{par}(z=0,r)$ becomes
independent
of the halo size $L$. Figure~\ref{fig:influence_l} displays this ratio
for various $L$ when formula (\ref{densite_rad}) is used.
The diffusion coefficient $K(E)$ and $l_{rad}$
are increasing functions of energy, so that when $l_{rad}$ becomes
of the order of the typical size of the halo ($L$ or $R$), diffusive
propagation
is affected by the shape and size of the halo.

We have presented here a toy model in which only one parent was
considered. Actually, there are several parents and the above conclusion
is not exact but remains approximate if they are considered,
so that at low energy, the ratio
$N^{uns}/\sum_{par} N^{par}$ is the same whatever the size of the halo.
All current data on radioactive nuclei are at low
energy, so that we do not expect such a ratio to give direct constraints
on $L$.

Nevertheless, the ratio of isotopes having the same parents, like
$^{10}$Be/$^9$Be, are usually considered.
As the quantity $^{10}$Be$/\sum_{par} N^{par}$ {\em does} depend on
the halo size
the ratio $^{10}$Be/$^9$Be eventually also does.



\subsection{Local sub-density: we all leave in a local bubble}
\label{sub-density}
Interest in the  study of our local environment has grown in the last
thirty years (see for example Cox et Reynolds (1987) for a
review).
As emphasized in McKee (1998), the very recent
all sky survey by {\sc rosat} became an invaluable resource
for all astronomers in general, and for the {\sc lism} study in
particular.
In this section, we present evidence indicating why it is
probably necessary to have a specific description for the local {\sc
ism}
(in a region of about $\sim$ 100 pc around the solar neighborhood).
We then develop a model  to incorporate the effect of these
local properties of the {\sc ism} on radioactive species within our
diffusion model.

\subsubsection{Main properties of the {\sc lism}}
\label{sec:lism}
First, it has to be noticed that far more progress has been made
in mapping
three dimensional distribution of galaxies many megaparsecs away
than the distribution of local interstellar
clouds within a hundred parsecs (McKee, 1998). We can nevertheless
summarize a few points about the local composition. First, the {\sc
lism} is defined
as a region of extremely hot gas ($\sim 10^5-10^6$ K) and low
density ($n\lesssim$ 0.005 cm$^{-3}$) within a bubble of radius between
$\lesssim$ 65-250 pc surrounded by a dense neutral gas boundary
(`hydrogen wall') (Sfeir et al., 1999; Linsky et al., 2000). A
smaller scale description of this bubble shows that the Sun
is located in a local fluff with  $N_{HI}\sim$ 0.1 cm$^{-3}$,
$T\sim 10^4$ K and a typical extension of $\sim 50$ pc.
It is of great importance for further modelling to realize that the
local
bubble is highly asymmetric (Cox et Reynolds, 1987;
Welsh et al.,  1994; Fruscione et al., 1994), and
that several cloudlets are present in the bubble (for a schematic
representation, see for example fig. 1. of Breitschwerdt et al., 2000).
Various models have attempted to explain the formation of this local
bubble (Smith and Cox, 2001; Breitschwerdt et al., 2000; see also
Cox, 1997 for a brief review), but this subject is far beyond our
concern.

Finally, a model can be built, considering
that the Sun is surrounded by a first shell
of radius $\sim 50$ pc and density $\sim$ 0.1 cm$^{-3}$,
and a second shell
of radius $\sim 200$ pc with an almost null density.
Beyond this second shell, we recover the usual average density
$1$ cm$^{-3}$ (or zero density if the radius of the second shell
extends beyond the disc).
In the rest of this paper, we use a coarser
description, in which the underdensity is modelled with one
cylindrical hole
of radius $r\un{hole}$ to be determined and with a null density.
It will be called `the hole' throughout this paper.

\subsubsection{Consequences for the radioactive production}

The propagation of cosmic rays is influenced by (i) the local magnetic
properties and (ii) the local matter content of the disc.
In this section, we investigate the effect of this hole on these
properties.

First, we assume that diffusion itself, as described by the
coefficient $K(E)$, is not affected by the presence of the hole,
{\em i.e.} diffusion is homogeneous.
Homogeneity seems to be confirmed by radio and
$\gamma$-ray observations, which can test {\em in situ} the spectrum
and density of cosmic rays (McKee 1998, Morfill \& Freyberg 1998).

The presence of the hole has another consequence.
Because the density is lower, there are less spallations
in the bubble, so that the secondary nuclei abundances (including
radioactive ones) are probably perturbed.
As a matter of fact, a realistic description of the matter content of
the
disc should take into account a spatial distribution (exponential or
more
complex) as used for example in Strong et Moskalenko (1998).
However, the only relevant quantity for stable nuclei is the  average
grammage of matter they cross. As these nuclei propagate in a
region which is much bigger than the hole, this average grammage is
only slightly changed by the presence of the hole.
This can be seen in figure~\ref{fig:dep_spatiale}, where the change in
the radial distribution of a stable species due to the hole is
represented.
In the following, the local density of stable nuclei will be computed
with a full matter disc of density $1\unit{cm}^{-3}$.

The situation is different for radioactive species, as
the typical distance they travel from their creation in the disc to
their detection on Earth is limited by their finite lifetime $\tau$.
This distance can be estimated as $l_{rad}\sim \sqrt{K(E) \gamma
\tau_0}$.
For the nuclei we consider here, and for energies of a few hundred
MeV/nuc,  $l_{rad}$ is of the same order of magnitude as the size of the
hole, so that their propagation is expected to be more strongly
perturbed.
To be specific, the fact that spallations do not occur within
the hole has three distinct physical effects.
First, it leads to a decrease in the spallation source term
of the radioactive species. Second, it also leads to a local decrease of
destructive spallations. Third, as there is less interstellar matter
to interact with, the energy losses are also lowered.

\section{Modelling the local propagation}

We now want to take these remarks and incorporate the
three physical effects discussed above in our diffusion model.
We start with a very simple model, in which the influence
of the hole can be easily understood.
We then adapt the model discussed at length in Paper I and
Donato et al. (2001) (hereafter Paper II).
In the following subsections, we present the
demonstration of the corresponding new formul\ae.
This section is a bit technical and the confident reader can skip the
demonstrations and go directly to subsection~\ref{subsec:conclusion1}.

\subsection{A first simple approach}
\label{App:hole_phys}

We begin with a very simple model, in which the galactic disc is an
infinitely thin disc extending to infinity, embedded in an infinite
diffusive volume. These assumptions are not unreasonable as propagation
of radioactive species is a local phenomenon and is not expected to be
much affected by boundaries.
We also ignore destructive spallations, galactic wind and energy losses.
At the origin of the galactic plane stands a hole of radius $r_{hole}$.

\begin{figure}[hbt!]
\centerline{\includegraphics[width=0.6\textwidth]{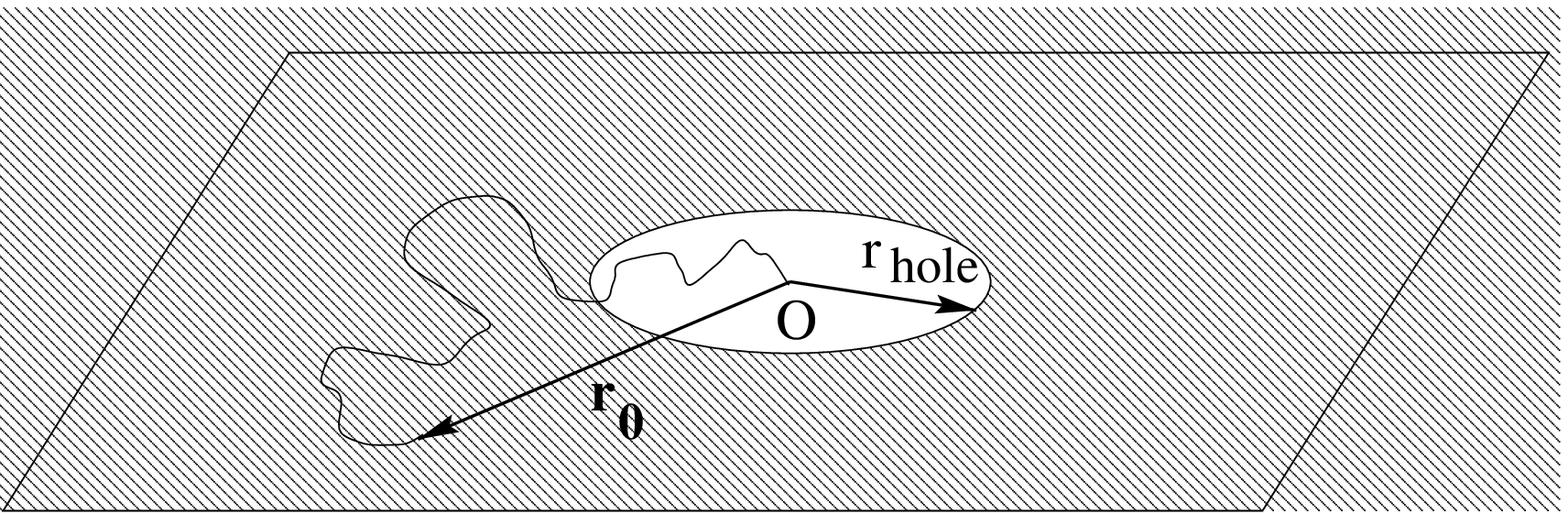}}
\caption{Schematic representation of the first model. The disc is
infinite in all directions and has zero thickness.}
\label{fig:schema1}
\end{figure}

Consider first the nuclei coming from spallations at time $t_0$ and at
a source point $\vec{r}_0$.
They act as an instantaneous source term $f_0 \delta(\vec{r}-\vec{r}_0)
\delta(t-t_0)$.
The number that would reach the center at time $t>t_0$, with no decay,
satisfies
\begin{displaymath}
         \frac{\partial f}{\partial t} =
         K \Delta f = \frac{K}{r^2} \frac{\partial}{\partial r} \left(
         r^2 \frac{\partial f}{\partial r} \right)
         =f_0 \delta(\vec{r}-\vec{r}_0) \delta(t-t_0)
\end{displaymath}
A solution of this equation is given by
\begin{displaymath}
         f(\vec{r}-\vec{r}_0, t-t_0) = \frac{f_0}{\left( 4\pi K
(t-t_0)\right)^{3/2}}
\exp \left\{  -
         \frac{\|\vec{r}-\vec{r}_0\|^2}{4 K (t-t_0)} \right\}
\end{displaymath}
For radioactive species with a lifetime $\tau = \gamma \tau_0$ (where
$\tau_0$ is the rest frame lifetime), decay must be taken into
account and we have
instead, at $\vec{r} = \vec{0}$,
\begin{equation}
         f(\vec{0},t) \propto \frac{1}{(t-t_0)^{3/2}} \exp \left\{ -
         \frac{r_0^2}{4 K (t-t_0)} \right\} \times \exp \left\{ -
         \frac{t-t_0}{\tau} \right\}
         \label{propagateur}
\end{equation}
This function is actually the propagator of this diffusion problem.
In the permanent regime, the total quantity of these nuclei at the
center of the hole is
obtained by summing the solutions for a distribution of point
sources acting at all instants from $t_0=-\infty$ to $t$ (now).
It leads to an integration over space and time
\begin{displaymath}
         N \propto \int_0^\infty 2\pi r_0 \, n(r_0) \, dr_0
\int_{-\infty}^t
         \frac{dt_0}{(t-t_0)^{3/2}} \exp \left\{ - \frac{r_0^2}{4 K
         (t-t_0)} \right\} \times \exp \left\{ -
         \frac{t-t_0}{\tau} \right\}
\end{displaymath}
in our case, $n(r_0)=0$ for $r_0<r_{hole}$ and $n(r_0)=n_0$ for
$r_0>r_{hole}$, so that
\begin{displaymath}
     N \propto n_0 \int_{r_{hole}}^\infty 2\pi r_0\, dr_0
     \int_{-\infty}^t dt_0 \,
         \frac{1}{(t-t_0)^{3/2}} \exp \left\{ - \frac{r_0^2}{4 K
(t-t_0)} \right\}
         \times \exp \left\{ - \frac{t-t_0}{\tau} \right\}
\end{displaymath}
Integration over $r_0$ is easily performed
\begin{displaymath}
         N \propto \int_{-\infty}^t  \frac{dt_0}{\sqrt{(t-t_0)}}
         \, \exp \left\{ - \frac{r_{hole}^2}{4 K (t-t_0)} \right\}
         \times \exp \left\{ - \frac{t-t_0}{\tau} \right\}
\end{displaymath}
Introducing $x = \sqrt{(t-t_0)/\tau}$,
\begin{displaymath}
         N \propto  \int_0^\infty dx \,
         \exp \left\{ - \frac{r_{hole}^2}{4 K \tau} \, x^{-2} -
x^2 \right\}
\end{displaymath}
This integral is given by
\begin{equation}
         \int_0^\infty dx \,
         \exp \left\{ - \frac{\alpha}{x^2} - x^2 \right\}
         = \frac{\sqrt{\pi}}{2} e^{-2 \sqrt{\alpha}}
         \label{integrale_a_la_con}
\end{equation}
so that finally
\begin{displaymath}
         N \propto e^{-r_{hole}/ \sqrt{K\tau}}
\end{displaymath}
The central density due to the full disc would be given by $r_{hole}=0$,
so that, introducing $l_{rad} \equiv \sqrt{K\gamma\tau_0}$,
\begin{equation}
         N_{r_{hole}}(r=0) = N_{(r_{hole}=0)}(r=0) \times
         e^{-r_{hole}/l_{rad}}
\end{equation}
We emphasize that in the problem treated here,
the time-dependent resolution by means of
propagators that we used in this section is equivalent to solving
the corresponding stationary  equation  ($\partial f / \partial t =
0$). This latter approach is more convenient for implementing the physical
effects we have neglected here. It will be used in the next sections and
as expected, the same behavior will be recovered.

\subsection{Analytical solutions without energy losses and
reacceleration}

We now turn to a more realistic model, namely a cylindrical diffusion
box of radius $R$ and half-height $L$, and a matter disc of half-height 
$h$, considered as infinitely thin. 
The local subdensity has to be represented by a circular
hole of radius $r_{hole}$ located at the position of the Sun.
Compared to the previous approach, it has the advantage of
correctly taking into account the presence of boundaries.
It also takes into account destructive spallations and galactic wind.
It is actually the model used in Paper I
and Paper II, with an additional hole of radius $r_{hole}$.

To make the problem tractable analytically, we make some simplifying
hypotheses.
First, to preserve the cylindrical symmetry, we consider
that the Sun (and the hole) is located at the center of the disc.
Second, we suppose that the density of the parent species is uniform
all over the galactic disc, {\em i.e.} it does not depend on the spatial
coordinate $r$.
The validity of these assumptions relies on the fact that the propagation
of radioactive species is a local process. We will discuss further
these points in section~\ref{subsec:verif_hypotheses}.

\begin{figure}[hbt!]
\centerline{\includegraphics[width=0.7\textwidth]{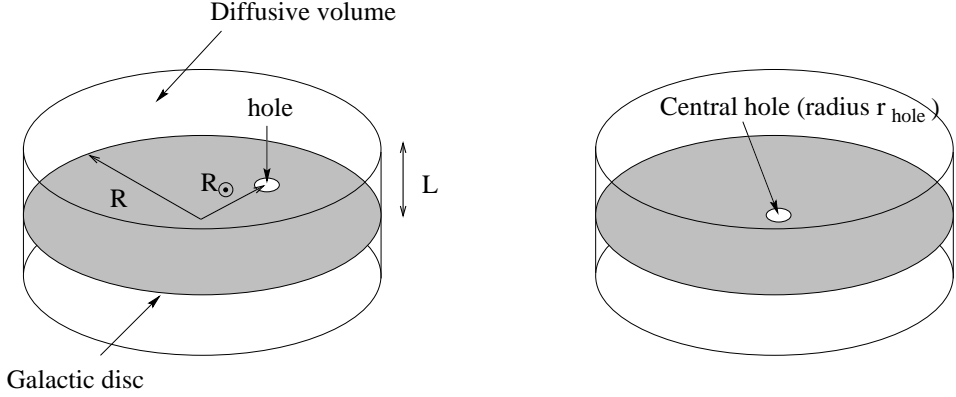}}
\caption{Schematic representation of the model. Left picture is the
actual geometry of the problem, with a hole of radius $r_{hole}$ in the
disc, at the position of the Sun. The right picture represents the
simplified geometry (cylindrical symmetry is preserved) which is
shown to give the same results.}
\label{fig:schemas}
\end{figure}

We want to solve the diffusion equation for a radioactive secondary
(no primary sources)
\begin{eqnarray}
      K \Delta N^{uns}(r,z) - V_c \frac{\partial N^{uns}(r,z)}{\partial
z}
      - \Gamma^{\beta} N^{uns}(r,z) + 2h \delta(z) \left[
      \Gamma^{prod}(r) N^{par}(r,0) - \Gamma^{inel}(r) N^{uns}(r,0)
      \right]=0
      \label{diffusion_equation}
\end{eqnarray}
To keep notations simple, we consider only one parent nucleus.
It would be straightforward to generalize to several parents.
Here, $N^{uns}$ denotes the unstable nucleus, $N^{par}$ the parent
nucleus
and $\Gamma^{prod}$, $\Gamma^{inel}$ and $\Gamma^{\beta}$ denote
respectively the production rate (by spallation) of the parent $N^{par}$
into the radioactive
nucleus $N^{uns}$, the destruction rate of $N^{uns}$ by inelastic
scattering and
its $\beta$--disintegration rate.
This equation differs from the no--hole case (equation (A1) of
Paper~I) because
the terms $\Gamma^{prod}(r)=n_{LISM}(r)v\sigma^{prod}$ and
$\Gamma^{inel}(r)=n_{LISM}(r)v\sigma^{inel}$ now depend
explicitly on $r$ {\em via} the local interstellar density which reads
\begin{equation}
n_{LISM}(r)=\Theta (r-r_{hole}) n_{ISM}
\end{equation}
where $\Theta$ is the Heavyside distribution.
Thus, we can rewrite equation (\ref{diffusion_equation}) in the form
\begin{equation}
      K \Delta N^{uns}(r,z) - V_c \frac{\partial N^{uns}(r,z)}{\partial
z}
      - \Gamma^{\beta} N^{uns}(r,z) + 2h \delta(z) \left[
      \Gamma^{prod}\Theta(r-r_{hole})N^{par}(r,0) -
\Gamma^{inel}\Theta(r-r_{hole})N^{uns}(r,0)
      \right]=0
      \label{diffusion_equation2}
\end{equation}
As for the no-hole model (see Paper I), a solution is found by
expanding the density over the Bessel functions $J_0(\zeta_i r/R)$
where the $\zeta_i$ are the zeros of $J_0$. The unknown quantities
$N^{uns}(r,z)$ are expanded as
\begin{equation}
        N^{uns}(r,z) = \sum_i N^{uns}_i(z) J_0 \left(\zeta_i \frac{r}{R}
\right)
        \label{deschamps}
\end{equation}
The known quantities are also expanded as
\begin{displaymath}
     \Upsilon(r,z) = \sum_i \Upsilon_i(z) J_0 \left(\zeta_i
     \frac{r}{R} \right)
     \;\;\; \mbox{with} \;\;\;
     \Upsilon_i(z) = \frac{2}{J_1^2(\zeta_i)} \int_0^1 \rho
\Upsilon(\rho,z)
     J_0(\zeta_i \rho) \, d\rho
\end{displaymath}
where $\rho=r/R$.

We now have to differentiate the case of spallation from that of
destruction.
For the first one, we can assume that the density of the parent
depends very smoothly on the
coordinate $r$ so that we can safely take
$N^{par}(r,0) = N^{par}(R_\odot,0)$; this approximation will be checked
in (\ref{subsec:verif_hypotheses}). We must compute the Bessel
transform of the Heavyside function
\begin{displaymath}
      \Theta_i =  \frac{2}{J_1(\zeta_i)^2} \int_0^1 \rho \,
      \Theta\left(\rho - \frac{r_{hole}}{R}\right) J_0(\zeta_i \rho)
d\rho
\end{displaymath}
With the notation $x=\zeta_i \rho$ and using the fact that a
primitive of $x J_0(x)$ is $x J_1(x)$,
it can be shown that
\begin{equation}
      \Theta_i =  \frac{2}{\zeta_i J_1(\zeta_i)^2} \times
      \left[ J_1(\zeta_i) - \frac{r_{hole}}{R} J_1\left(\zeta_i
\frac{r_{hole}}{R}
      \right) \right]
      \label{fio}
\end{equation}

For the destructive spallation term, we have to Bessel develop the
function $\Theta(r-r_{hole})N^{uns}(r,0)$. At variance with the
precedent case, the distribution $N^{uns}(r,0)$ is expected to vary significantly
across the hole so that the previous approximation cannot be made.
We thus have
\begin{displaymath}
    \Theta(r-r_{hole})N^{uns}(r,0) = \sum_{i=1}^\infty \Omega^{uns}_i
    J_0 \left(\zeta_i \frac{r}{R} \right)
\end{displaymath}
where
\begin{displaymath}
     \Omega^{uns}_i \equiv \frac{2}{J_1^2(\zeta_i)}
     \int_0^1 \rho \, N^{uns}(\rho,0) \Theta \left(\rho -
     \frac{r_{hole}}{R} \right) J_0(\zeta_i
     \rho) \, d\rho
\end{displaymath}
Inserting the Bessel development of $N^{uns}(r,z=0)$ into this relation
(see equation~\ref{deschamps}), we find
\begin{displaymath}
     \Omega^{uns}_i = \frac{2}{J_1^2(\zeta_i)}  \sum_{j=1}^\infty
     N^{uns}_j(z=0) \int_{r_{hole}/R}^1 \rho \,
     J_0(\zeta_j \rho) J_0(\zeta_i \rho) \, d\rho
\end{displaymath}
Using the property
\begin{displaymath}
        \int \rho \, J_0(\zeta_j \rho) J_0(\zeta_i \rho) \, d\rho
        = \left\{ \begin{array}{ll} \displaystyle
        \frac{1}{\zeta_j^2 - \zeta_i^2} \left[
        - \zeta_i \rho J_0(\zeta_j \rho) J_1(\zeta_i \rho)
        + \zeta_j \rho J_1(\zeta_j \rho) J_0(\zeta_i \rho) \right]
        & \mbox{for $i\neq j$}\vspace{5mm} \\ \displaystyle
        \nonumber \int \rho \, J_0^2(\zeta_i \rho) \, d\rho
        = \frac{1}{2} \rho^2 \left[
        J_0^2(\zeta_i \rho) + J_1^2(\zeta_i \rho) \right]
        & \mbox{else} \end{array} \right.
\end{displaymath}
it follows that
\begin{eqnarray}
        \Omega^{uns}_i &=& N^{uns}_i(0)
        - \frac{r_{hole}^2}{R^2 J_1^2(\zeta_i)} N^{uns}_i(0)
        \left[ J_0^2\left(\frac{\zeta_i r_{hole}}{R}\right)
        + J_1^2\left(\frac{\zeta_i r_{hole}}{R}\right)\right]
\label{Omega_i}\\
        &+& \frac{2r_{hole}}{RJ_1^2(\zeta_i)} \sum_{j\neq i}
        \frac{N^{uns}_j(0)}{\zeta_j^2 - \zeta_i^2} \left[
        \zeta_i J_0\left(\frac{\zeta_j r_{hole}}{R}\right)
        J_1\left(\frac{\zeta_i r_{hole}}{R}\right)
        - \zeta_j J_1\left(\frac{\zeta_j r_{hole}}{R}\right)
        J_0\left(\frac{\zeta_i r_{hole}}{R}\right) \right] \nonumber
\end{eqnarray}
Finally, putting everything altogether,
the Bessel transform of equation~(\ref{diffusion_equation2})
reads
\begin{equation}
     \left[ \frac{\partial^2}{\partial z^2} - \frac{V_c}{K}
     \frac{\partial}{\partial z} - \left( \frac{\zeta_i^2}{R^2} +
     \frac{\Gamma^\beta}{K} \right) \right] N^{uns}_i(0)
     + \frac{2h}{K} \delta(z) \left[
     \Theta_i \times \Gamma^{prod} N^{par}(R_\odot,0)
     - \Omega^{uns}_i \times \Gamma^{inel} \right] = 0
     \label{eq:bessel_final}
\end{equation}
where $\Theta_i$ and $\Omega_i^{uns}$ are given by equations~(\ref{fio})
and~(\ref{Omega_i}).
This equation is now very similar to equation~(A6) of Paper I, and
resolution proceeds as exposed therein.
The solution in the halo is
\begin{displaymath}
      N^{uns}_i(z) \propto e^{V_c z/2K} \times \mbox{sinh} \left\{
      \frac{S_i (L-z)}{2} \right\}
\end{displaymath}
where $S_i$ and $A_i$ were defined in section~(\ref{halo_size}).
The solution in the disc is obtained by integrating equation
(\ref{diffusion_equation}) across the disc,
\begin{displaymath}
        2{N_i^{uns}}'(0) - 2 N^{uns}_i(0) \frac{V_c}{K} - \frac{2h
        \Gamma^{inel}}{K} \Omega^{uns}_i + \frac{2h
        \Gamma^{prod}}{K} \Theta_i N^{par}(R_{\odot},0) = 0
\end{displaymath}
which gives
\begin{equation}
      N^{uns}_i(0)  = \Theta_i \times \frac{2h \Gamma^{prod}
      N^{par}(R_{\odot},0) }{A_i} - \frac{2h \Gamma^{inel}}{A_i}
      \left[ \Omega^{uns}_i - N^{uns}_i \right]
      \label{eq:bessel3}
\end{equation}
The first term is very similar to the secondary source contribution
of a no-hole model, but with an additional $\Theta_i$ factor taking
correctly into account the effect of the hole on production
spallations.
The second term takes into account the effect of the hole on
destructive spallations. It is expressed as
\begin{eqnarray}
     \Omega^{uns}_i - N^{uns}_i  &=&
     - \frac{r_{hole}^2}{R^2 J_1^2(\zeta_i)} N^{uns}_i(0)
     \left[ J_0^2\left(\frac{\zeta_i r_{hole}}{R}\right)
     + J_1^2\left(\frac{\zeta_i r_{hole}}{R}\right)\right]
     \label{bob}\\
     &+& \frac{2r_{hole}}{RJ_1^2(\zeta_i)} \sum_{j\neq i}
     \frac{N^{uns}_j(0)}{\zeta_j^2 - \zeta_i^2} \left[
     \zeta_i J_0\left(\frac{\zeta_j r_{hole}}{R}\right)
     J_1\left(\frac{\zeta_i r_{hole}}{R}\right)
     - \zeta_j J_1\left(\frac{\zeta_j r_{hole}}{R}\right)
      J_0\left(\frac{\zeta_i r_{hole}}{R}\right) \right] \nonumber
\end{eqnarray}
As this term depends on $N^{uns}_j(0)$,  (\ref{eq:bessel3})
is a coupled set of equations and may be tricky to solve in practice.

Actually, as the effect of spallations is expected to be small, we
use a perturbative method.
We first compute the solution of equation (\ref{eq:bessel3})
without the last term of equation~(\ref{bob}),
\begin{displaymath}
      N^{uns, (0)}_i(0)  = \Theta_i \times \frac{2h \Gamma^{prod}
      N^{par}(R_{\odot},0) }{A_i'}
\end{displaymath}
with
\begin{displaymath}
      A_i'  = A_i + 2h \Gamma^{inel} \frac{r_{hole}^2}{R^2
J_1^2(\zeta_i)}
      \left[ J_0^2\left(\frac{\zeta_i r_{hole}}{R}\right)
      + J_1^2\left(\frac{\zeta_i r_{hole}}{R}\right)\right]
\end{displaymath}
The term we have neglected can now be taken into account by
substituting the $N^{uns}_j(0)$ in
equation~(\ref{bob}) by
these zero-order solutions $N_j^{uns, (0)}$.
\begin{displaymath}
        N^{uns, (1)}_i(0) =  N^{uns, (0)}_i(0) +
        \frac{4h\Gamma^{inel}}{J_1^2(\zeta_i) A_i}
        \, \frac{r_{hole}}{R} \times \sum_{j\neq i}
        \frac{N^{uns, (0)}_j(0)}{\zeta_j^2 - \zeta_i^2}
        \left[\zeta_i J_0\left(\frac{\zeta_j r_{hole}}{R}\right)
        J_1\left(\frac{\zeta_i r_{hole}}{R}\right)
        - \zeta_j J_1\left(\frac{\zeta_j r_{hole}}{R}\right)
        J_0\left(\frac{\zeta_i r_{hole}}{R}\right) \right]
\end{displaymath}
The new quantities $N^{uns, (1)}_i(0)$ can then be used instead of
$N^{uns, (0)}_i(0)$ to estimate the correction at the next order.
This procedure is repeated until convergence is reached.
In practice, convergence is very quick and we only need to iterate
twice.

\subsection{Behavior of the solution as a function of the hole radius}
\label{subsec:behavior}

\begin{figure}[hbt!]
\centerline{\includegraphics[width=0.47\textwidth]{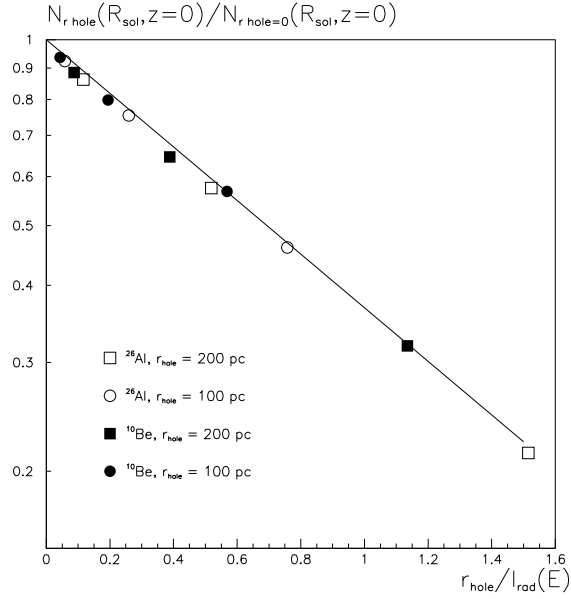}}
\caption{The radioactive flux is computed for
$^{26}$Al and $^{10}$Be, for several energies (corresponding to
several values of $l_{rad}(E)$) and for several hole
radii $r_{hole}$. This plot displays each flux divided
by the flux obtained in the homogeneous case ($r_{hole}=0$), as a
function of $r_{hole}/l_{rad}$.
The solid line represents the function $\exp(-r_{hole}/l_{rad})$.}
\label{fig:dep_a}
\end{figure}

A numerical study of this solution (see figure~\ref{fig:dep_a})
shows that the dependence in $r_{hole}$
can be expressed as, with good accuracy,
\begin{equation}
        N_i(0) \approx \exp \left( -\frac{r_{hole}}{l_{rad}} \right)
        \;\;\; \mbox{with} \;\;\;
        l_{rad}=\sqrt{K(E)\gamma\tau_0}
        \label{dep_a}
\end{equation}
\begin{table}[ht]
    \caption{Values of $l_{rad}$ for several radioactive nuclei and energies.}
    \label{valeurs_l_rad}
    \begin{center}
	\begin{tabular}{|c|c|ccc|}   \hline
	    & $\tau_0$ (Myr) & $E_k=100 \unit{MeV/n}$ & $1
	    \unit{GeV/nuc}$ & $10 \unit{GeV/nuc}$ \\ \hline
	    $^{10}$Be & 2.17 & $0.075 \unit{kpc}$ & $0.22\unit{kpc}$ &  $0.95
	    \unit{kpc}$\\
	    $^{26}$Al & 1.31  & $0.037\unit{kpc}$ & $0.11\unit{kpc}$ &
	    $0.47\unit{kpc}$\\
	    $^{36}$Cl & 0.443 & $0.019\unit{kpc}$ & $0.056\unit{kpc}$ &
	    $0.25\unit{kpc}$ \\
	    \hline
	\end{tabular}
    \end{center}
\end{table}
The physical meaning of this dependence is better seen with the first
approach (section~\ref{App:hole_phys}).
The relevant quantities to compare
are the size of the hole and $l_{rad}$, which represents the typical
distance
travelled by a radioactive nucleus before it decays. This distance
increases with energy, because of time dilation and because
high energy nuclei diffuse more efficiently ($K$ grows with energy).
For a diffusion coefficient of the form $K=K_0 \beta {\cal R}^\delta$,
$l_{rad}$ can be expressed as
\begin{equation}
       l_{rad} = \sqrt{K \gamma \tau_0} = \sqrt{K_0 \tau_0}\,
       \frac{A^{(\delta-1)/4}}{Z^{\delta/2}} \left( E_{k,nuc}^2 + 2m_p
       E_{k,nuc} \right)^{(\delta+1)/4}
       \label{l_rad}
\end{equation}
Typical values are given at several energies and for several radioactive
nuclei for $K_0 = 0.033 \unit{kpc}^2 \unit{Myr}^{-1}$
and $\delta=0.6$ (a good model taken from Paper I) in Table
$\ref{valeurs_l_rad}$.

\subsection{Discussion of the validity of the hypotheses}
\label{subsec:verif_hypotheses}
\begin{figure}[ht]
\centerline{\includegraphics[width=.5\textwidth]{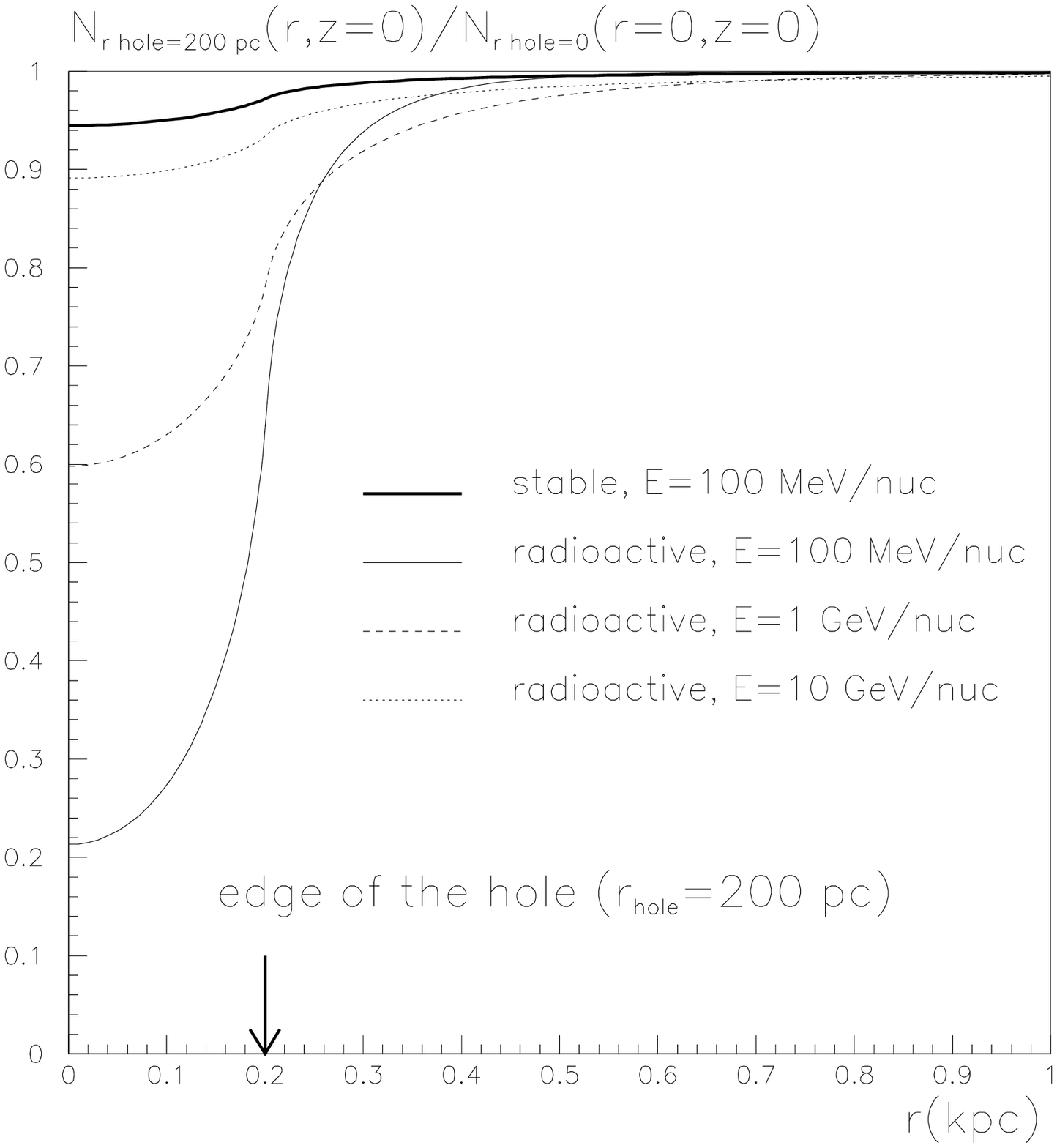}}
\caption{Radial distribution of a radioactive species in the disc,
across
the hole, for $r_{hole}=200 \unit{pc}$. Numerical values were taken for
$^{26}$Al (radioactive) and $^{28}$Si (stable).
The distribution of radioactive is very sensitive to the presence of
the hole. However this effect is local and vanishes at several
$r_{hole}$.
On the contrary, the distribution of stable species is not much
affected.}
\label{fig:dep_spatiale}
\end{figure}

The most {\em a priori} questionable assumption is that we located the
hole at the center of the galactic disc, though we know that the Sun
is at a galactocentric distance of about $R=8 \unit{kpc}$.
Figure~\ref{fig:dep_spatiale} displays the radial distribution of a
radioactive species in the disc for several energies. We see this distribution is only
perturbed locally by the presence of the hole: at a distance of
a few $l_{rad}$, the flux is no longer affected by the hole.
Conversely, the flux at the center of the hole is not affected by the
boundaries of the diffusive box if they are farther than a few
$l_{rad}$. We could even consider that the galactic disc is an
infinite plane. This means that the same distribution -- and the same
reduction factor $\exp(-r_{hole}/l_{rad})$ -- is obtained whatever the
position of the hole, as long as it is far enough from the edges of
the box (compared to $l_{rad}$), which is the case for us.
The density of parent nuclei must obviously be estimated at the
position of the Sun (and not at the center of the galaxy !).

We also assumed that the parent has a homogeneous distribution all
over the disc, which is known to be wrong: the parent flux is greater
around the position of the sources ($R \sim 4 \unit{kpc}$, see for
example
Paper I, section 3.5.1 and references therein)
and decreases regularly to zero at the edge of the disc ($R = 20
\unit{kpc}$). However, as the radioactive propagation is a local
phenomenon, at the scale of $l_{rad}$, it is a good approximation as
long as the parent distribution does not vary much on the spatial
scale of $l_{rad}$. Actually, even if the parent distribution varies
linearly with radius, the approximation remains good.
The probability $d{\cal P}$ that a radioactive nucleus detected at
the center of
the hole has been created by a spallation at a distance between $r$
and $r+dr$ can be
computed explicitly in the simple model described in
subsection~\ref{App:hole_phys}, and it is found that (see Appendix for
the demonstration)
\begin{equation}
    d{\cal P} (\mbox{emitted between $r$ and $r+dr$} |
    \mbox{detected at center} )
    \propto \Theta(r-r_{hole}) \, N_{parent}(r) \, e^{-r/l_{rad}} \, dr
    \label{proba_inverse}
\end{equation}
so that most of the radioactive nuclei come from a ring extending from
the edge of the hole ($r=r_{hole}$) to a few $l_{rad}$ away.
As a result, if the parent nucleus density is not uniform,
the spallation rate is determined by the effective density obtained as
the average over the disc with the appropriate weight, given in the
expression~\ref{proba_inverse2} (see Appendix)
\begin{displaymath}
    N_{parent}^{ef\!fective} =
    \frac{\int \!\!\! \int d^2\vec{r} \, N_{parent}^2(\vec{r}) \,
    \Theta(r-r_{hole}) \, \exp(-r/l_{rad})}
    {\int \!\!\! \int  d^2\vec{r}\, N_{parent}(\vec{r}) \,
    \Theta(r-r_{hole}) \, \exp(-r/l_{rad})}
\end{displaymath}
We have compared this effective parent density to the actual parent
density
$N_{parent}(R_{\odot}, z=0)$
at the location of the Sun. For hole radii $r_{hole} < 400 \unit{pc}$
and for $l_{rad}< 5 \unit{kpc}$, the difference is less than one
percent at 10 GeV/nuc, and is completely negligible for $E \lesssim 1
\unit{GeV/nuc}$.

\subsection{Energy losses and reacceleration}
In homogeneous models, the energy changes induced by energy losses and 
reacceleration are described by the equation (see Paper I and Paper II)
\begin{equation}
       A_i N_i + 2h \frac{\partial}{\partial E} \left\{
       b(E) N_i(E)- K_{EE} \frac{\partial N_i}{\partial E} \right\}
       = A_i N_i^{(0)}
       \label{pertes}
\end{equation}
where $N_i$ is the Bessel transform of the final flux and $N_i^{(0)}$
is the initial flux (before energy losses and reacceleration are
applied).
Because of the hole, the energy loss term $b(E)$ now has a spatial
dependence.
It has the same value as above, except in the hole where
there is no matter to interact with, so that
\begin{displaymath}
      b(r,E) = \Theta(r-r_{hole}) b(E)
\end{displaymath}
Thus, the Bessel transform of the quantity $\Theta(r-r_{hole}) N(r,E)$,
denoted $\mho_i$, must be introduced.
As the spatial dependence of this function is exactly the
same as the spallation term of the previous section, it is
straightforward to write (see eq. \ref{Omega_i})
\begin{eqnarray}
       \mho_i(E) &=& N_i(E) \left\{ 1 +
       \frac{r_{hole}^2}{R^2 J_1^2(\zeta_i)}
       \left[ J_0^2 \left(\frac{\zeta_i r_{hole}}{R}\right)
       + J_1^2 \left(\frac{\zeta_i r_{hole}}{R}\right) \right] \right\}\\
       &-& \frac{2r_{hole}}{RJ_1^2(\zeta_i)} \sum_{j\neq i}
       \frac{N_j}{\zeta_j^2 - \zeta_i^2} \left[
       - \zeta_i J_0 \left(\frac{\zeta_j r_{hole}}{R}\right)J_1
       \left(\frac{\zeta_i r_{hole}}{R}\right)
       + \zeta_j J_1 \left(\frac{\zeta_j r_{hole}}{R}\right)J_0
       \left(\frac{\zeta_i r_{hole}}{R}\right)\right] \nonumber 
       \;\;\; ,
\end{eqnarray}
so that $b(E)N_i(E)$ in (\ref{pertes}) has to be replaced by
$b(E)\mho_i(E)$.
It is again a coupled set of equations. The same perturbative approach
as before is used: the new quantities $N_i^{(1)}$ affected by energy
losses are
estimated by replacing $N_i$ by $N_i^{(0)}$. The new set $N_i^{(1)}$
is then inserted instead of $N_i^{(0)}$, etc\ldots until convergence
is reached.
It is found that even in the presence of energy losses and
reacceleration, the dependence
on the hole radius is still very well described by (\ref{dep_a}).

We note that reacceleration has a negligible effect on radioactive
propagation, which may be understood as it takes much longer than
the lifetime $\gamma \tau_0$ to significantly reaccelerate a nucleus.

\subsection{Conclusion and inclusion in the propagation code}
\label{subsec:conclusion1}

The presence of a hole of radius $r_{hole}$ lowers the radioactive flux
by a factor given with a good precision by $\exp(-r_{hole}/l_{rad})$ 
where $l_{rad}$ is given in (\ref{l_rad}).

It is not possible in practice to directly use the analytical
formul\ae \ described above because the
double sums over Bessel indices are far too time consuming.
Indeed, the accurate description of the hole requires a development
of all functions over at least the $N \gg R/r_{hole} \pi \sim 100$ first
functions $J_0(\zeta_i x)$. We used $N=5000$, for which a very good
convergence of the Bessel series was achieved, and the double
summation takes $N^2 = 2.5 \times 10^7$ elemental steps for each
iteration and each model. The computation time required to apply
these formul\ae\  to all the models would have been too large.

Thus, we preferred to take advantage of the exponential dependence
discussed above, and
the flux of a given radioactive nucleus in the presence of a hole is
obtained from each model with no hole by a multiplication by
$\exp(-r_{hole}/l_{rad})$.


\section{Experimental data and configurations of parameter space}
We now turn to the more standard diffusion aspects of propagation.
The main ingredients of our diffusion model have been widely
depicted in Paper I, and we will not be exhaustive about them here.
Instead, we simply motivate the choice of configurations that are
used in this paper. We also review the experimental data that
we compared to our calculations.

\subsection{Sets of configurations used for the analysis}
\label{parameter_space}

Propagation is assumed to be a diffusive process
occurring both in the galactic disc and in a halo of thickness $L$.
This process is characterized by an energy--dependent diffusion coefficient
of the form $K(E)= K_0 \beta {\cal R}^\delta$ where $K_0$ and $\delta$
are parameters of the model and ${\cal R}$ is the rigidity.
Propagation is affected by a galactic wind $V_c$ perpendicular to the
galactic plane and by the presence of Alfv\'en waves of velocity $V_a$.
These are the five parameters of the model.
All the sets of parameters consistent with measurements of stable nuclei

analysis have been extracted and discussed in Paper I (see
this paper for a detailed presentation of the model and more
specifically figures 7 and 8).
These sets of parameters are also consistent with the antiproton
spectrum, as shown in Paper II.

Some further cuts in our initial sets of parameters could probably be
applied on physical grounds (see for example the discussion on the
Alfv\'enic velocity $V_a$ in section 5.2, Paper II).
We adopted a conservative attitude and used the whole set,
as the aim of this paper is precisely to explore whether
it is possible to obtain further constraints on these parameters
from the study of radioactive species.
\subsection{The data set we used}
\label{sec:dataset}
Several experiments in the last twenty or thirty years have measured
radioactive isotopes in cosmic rays with increasing precision,
at energies of a few hundred MeV/nuc.
The early data -- usually presented as the ratio of the radioactive
isotope to its stable companion(s) -- were affected by errors of around
25--30$\%$.
The latest published data have error bars reduced
by a factor of two or three.
In the following, we implicitly refer to
three satellite experiments, namely Voyager, Ulysses and {\sc ace}.
Other experiments will sometimes be shown on figures
but they will be purely illustrative since their accuracy is far
smaller.

The  best measured ratio is probably $^{10}{\rm
Be}/^{9}{\rm Be}$ which corresponds to the lowest Z
$\beta$--radioactive nucleus.
Data from Ulysses (Connell, 1998) and from {\sc ace} (Binns et al.,
1999) are
consistent, the quoted error bars being smaller for {\sc ace}.
They are also consistent with the Voyager data point (Lukasiak et al.,
1999) for which the quoted error is larger.
We do not use the {\sc smili} data, as the possibility that they are
due to a statistical fluctuation is not ruled out (Ahlen et al., 2000).

As regards the radioactive chlorine isotope $^{36}$Cl,
results are usually
provided as $^{36}$Cl to total Cl ratio. The only available data, to
our knowledge, are those from Ulysses (Connell et al., 1998) with a
1$\sigma$
error of about 35\%, and {\sc ace} (Binns et al., 1999)
whose errors (even taken at 3$\sigma$) are completely included in the
Ulysses 1-$\sigma$ upper error band.

We finally end with the $^{26}{\rm Al}/^{27}{\rm Al}$ ratio.
Contrary to other radioactive ratios, the measurements show more problems.
Indeed, 1$\sigma$ data from Ulysses (Simpson and Connell, 1998)
and {\sc ace} (Binns et al., 1999) exclude each other (the {\sc ace} central 
point is much lower than Ulysses').
Enlarging {\sc ace} error bars (which are smaller than Ulysses) to 3$\sigma$
does not improve the compatibility. On the other hand, Ulysses data are
fully compatible with 1$\sigma$ Voyager data (Lukasiak et al., 1994),
whose uncertainty is still much greater than for the other two experiments.
The possible discrepancy between some of these data will be
addressed later.


\section{Results}

We now present our analysis of the radioactive nuclei
abundances. First, in the diffusion model with no hole (hereafter
the {\em homogeneous model}),
we can compute the expected flux of radioactive nuclei for each set of
diffusion parameters given by the analysis of stable nuclei (Paper I)
and compare to the data.
Then, the presence of a local underdense region discussed above is
tested, and the radius $r_{hole}$ of this region is introduced as an
additional parameter.
The corresponding models will be called {\em inhomogeneous models}.
As the stable nuclei are almost not affected by the presence of the
hole (see figure~\ref{fig:dep_spatiale}),
the sets of diffusion parameters given in Paper I are also used
in the inhomogeneous case.

To sum up the procedure, the sets of diffusion parameters ($K_0$, $L$,
$\delta$, $V_a$ and $V_c$) given in Paper I are used to compute the
radioactive nuclei fluxes, for different hole radii $r_{hole}$. 
The special case $r_{hole}=0$ corresponds to the homogeneous models.

We proceed as follows.
We first focus independently on the $^{10}$Be/$^9$Be and $^{36}$Cl/Cl
ratios.
They are computed for all the sets of diffusion parameters compatible
with B/C (given in Paper I) and for several hole radii $r_{hole}$.
The constraints they induce on the parameters are studied.
We then analyze simultaneously the two ratios $^{10}$Be/$^9$Be and
$^{36}$Cl/Cl
and finally the three ratios $^{10}$Be/$^9$Be, $^{36}$Cl/Cl and
$^{26}$Al/$^{27}$Al.

\subsection{Constraints from $^{10}$Be/$^9$Be}
\subsubsection{Comparison to {\sc ace} data}
To begin with, we consider only the $^{10}$Be/$^9$Be ratio,
for which we have three compatible measurements.
This is also the ratio for which an accurate spectrum is likely to be
available  in a near future (by {\sc isomax}, {\sc ams}, \ldots).
This ratio is computed for each of the 5-parameter sets compatible
with B/C given in Paper I, and for several hole radii $r_{hole}$.
The result is compared to the {\sc ace} data and the parameters giving a
ratio
falling out of the 3$\sigma$ {\sc ace} error bars are discarded.
\begin{figure}[hbt!]
\centerline{\includegraphics*[width=0.67\textwidth]{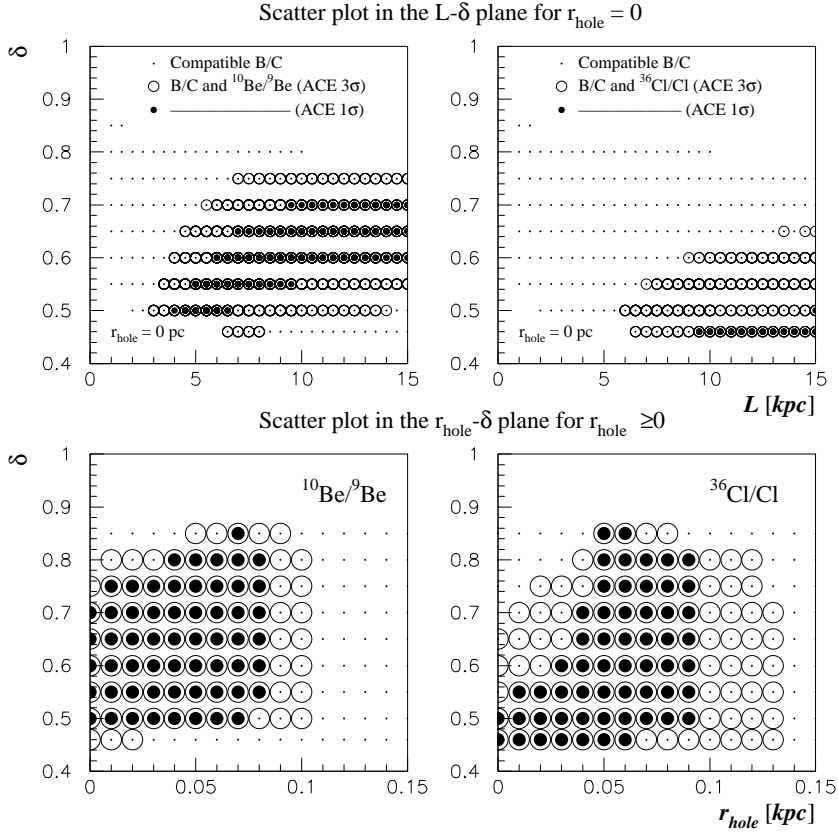}}
\caption{Representation of the models compatible with
combination of B/C (see Paper I) and various radioactive {\sc ace} data.
See text and notations in the figure. Upper panels display
homogeneous models ($r_{hole}=0$) in the plane $L$--$\delta$.
Lower panels display inhomogeneous models ($r_{hole}\geq 0$) in the
plane $r_{hole}$--$\delta$.}
\label{fig:be_cl}
\end{figure}
We also do the same with a more stringent limit of 1$\sigma$,
and the results are shown in the left panels of figure~\ref{fig:be_cl},
where the scatter of the configurations is plotted in
the plane $\delta$--L for the homogenous case, and $\delta$--$r_{hole}$
in the inhomogeneous one. For illustrative purpose, we
also show on the right panels of figure~\ref{fig:be_cl} the result of
the same procedure with 3$\sigma$ and 1$\sigma$ $^{36}$Cl/Cl {\sc ace}
data.
The dots in this figure are merely the models obtained
in the Paper I analysis.

The upper figure shows that  for homogeneous models, the
$^{10}$Be/$^9$Be ratio alone further constrains $L$
to a smaller region of  the parameter space.
For inhomogeneous models (lower part),
the hole radius is constrained to
values $r_{hole} \lesssim 100 \unit{pc}$.
This can be easily understood:
the value of $^{10}$Be/$^9$Be in an inhomogeneous model is given by the
corresponding value in a homogeneous model, decreased by the
exponential factor $\exp(-r_{hole}/l_{rad})$.
As the initial homogeneous parameter sets
compatible with B/C give a wide range of $^{10}$Be/$^9$Be values,
those which are larger than the data are redeemed in inhomogeneous
models.
For holes larger than $100 \unit{pc}$, the exponential decrease is too
important and theoretical predictions are too low to fit {\sc ace} data.

If we do not believe in the presence of the bubble, then our
conclusions are similar to previous works: higher values of $L$ are
preferred in homogeneous models ($L \gtrsim 4 \unit{kpc}$).
For inhomogeneous models (including $r_{hole}=0$), the allowed ranges
for the other diffusion parameters are not much affected if compared
to the results of the stable nuclei analysis (Paper I).
The $^{36}$Cl/Cl ratio yields similar conclusions.

\subsubsection{Discussion}

Here we want to discuss qualitatively two other important points:
the spectral behaviour and the interrelation of the three radioactive
ratios.
It will allow a more intuitive grasp of the combined analysis presented
in the next section.

\begin{figure}[hbt!]
\centerline{\includegraphics*[width=0.79\textwidth]{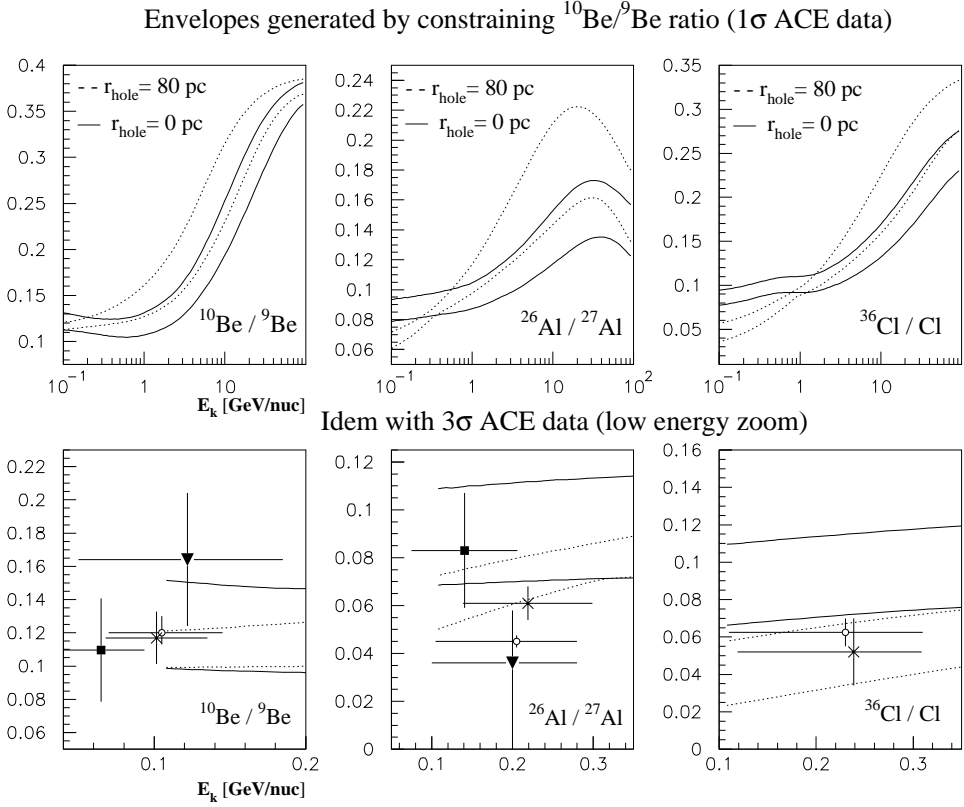}}
\caption{Envelopes of the spectra obtained with all the models
compatible with $^{10}$Be/$^9$Be {\sc ace} 1$\sigma$ (upper) and
3$\sigma$ (lower) for the three radioactive species.
Solid lines are for homogeneous models ($r_{hole}=0$ pc) and dashed
lines
are for inhomogeneous models ($r_{hole} = 80$ pc).
Data in the lower panel are from {\sc ace} (circles), Ulysses
(crosses), Voyager (filled squares) and {\sc isee} (filled
triangles, Wiedenbeck, 1985)}
\label{fig:envelopes}
\end{figure}

We first select all the homogeneous models compatible with
$^{10}$Be/$^9$Be (1$\sigma$ and 3$\sigma$ {\sc ace} data).
Then, for each point of the resulting subset, the full
propagation code is run and for the three radioactive species,
maximal and minimal ratios are calculated at all energies and
generate an envelope.
These  envelopes are represented in the upper part of
figure~\ref{fig:envelopes}.
They contain all the allowed values of the above-mentioned ratios which
are "compatible with B/C" and "compatible with $^{10}$Be/$^9$Be
{\sc ace} data $1\sigma$".
The same procedure is applied to the inhomogeneous models
$r_{hole}= 80 \unit{pc}$ (this value will be favored in the combined
analysis).

We now want to address the following questions: (i) is it possible,
from the radioactive spectra, to experimentally distinguish
inhomogeneous
from  homogeneous models, (ii) is it possible to break the degeneracy
in the propagation parameters, and (iii) what would be a clear
signature of the presence of the hole ?

As regards the first question, the upper left panel
($^{10}$Be/$^9$Be) clearly shows that the
answer is no, as even for extreme models ($r_{hole} \sim 80
\unit{pc}$), there is still an overlap between the homogeneous and
inhomogeneous envelopes.
Concerning question (ii), a partial answer has been given in the
previous
section: the $^{10}$Be/$^9$Be ratio is degenerate in
the diffusion parameters at the {\sc ace} energy.
Figure~\ref{fig:envelopes} shows that it is actually true for all
energies.
We finally turn to question (iii).
From left to right panels of figure~\ref{fig:envelopes}, the nuclei
have smaller and smaller lifetimes, so that the influence of the hole is
larger and larger.
In particular, the ratio $^{36}$Cl/Cl is the most sensitive
to $r_{hole}$, and at low energy, homogeneous and inhomogeneous
envelopes (for $r_{hole}=80 \unit{pc}$) do not overlap.
We thus expect that the combination of these ratio may break the
degeneracy in $r_{hole}$.

We also show, in the lower part of figure~\ref{fig:envelopes}, the
envelopes obtained as above but with the weaker constraint that
$^{10}$Be/$^9$Be
is in the 3$\sigma$ {\sc ace} error bars.
For $r_{hole}= 80 \unit{pc}$ and at the {\sc ace}
data energy, the $^{10}$Be/$^9$Be is never larger than $0.12$, which
corresponds to the largest $^{10}$Be/$^9$Be obtained in the
homogeneous models, decreased by the corresponding exponential factor.
This is not shown in the figures, but for $r_{hole}= 50 \unit{pc}$,
the lower panel dotted lines would be
almost superimposed on the solid lines, whereas for $r_{hole}= 100
\unit{pc}$, they would only delimitate a very narrow strip.
We also see that inhomogeneous models ($r_{hole}=80 \unit{pc}$)
are favored by the  $^{36}$Cl/Cl data.
On the other hand, all the models compatible with $^{10}$Be/$^9$Be
clearly fail to reproduce the 3$\sigma$ {\sc ace} $^{26}$Al/$^{27}$Al
data.

\subsection{Combined analysis of $^{10}$Be/$^9$Be and $^{36}$Cl/Cl}

\begin{figure}[b]
\centerline{\includegraphics*[width=0.8\textwidth]{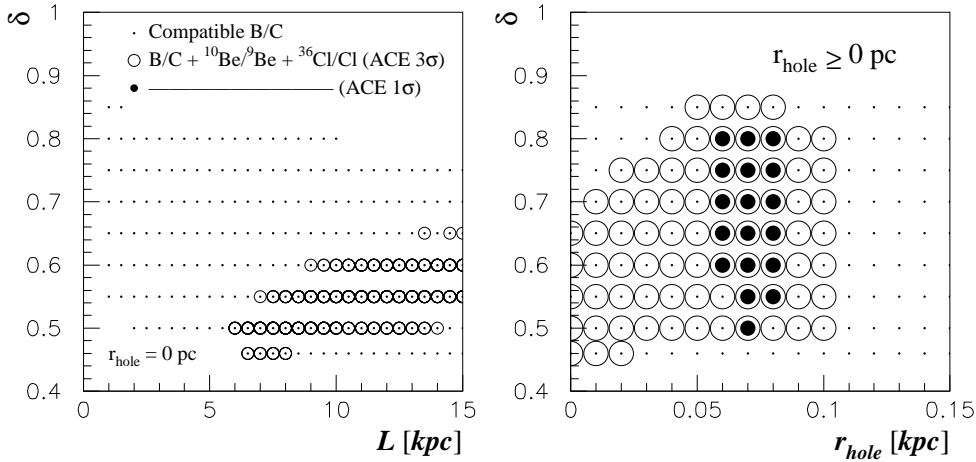}}
\caption{Representation of the models compatible with B/C plus both
$^{10}$Be/$^9$Be
and $^{36}$Cl/Cl {\sc ace} 3$\sigma$ (open circles) and 1$\sigma$
(filled circles).
Left panel displays homogeneous models ($r_{hole}=0$) in the plane
$L$--$\delta$.
Right panel displays inhomogeneous models ($r_{hole}\geq 0$) in the
plane
$r_{hole}$--$\delta$.}
\label{fig:be_et_cl}
\end{figure}

We now analyze simultaneously the two ratios $^{10}$Be/$^9$Be and
$^{36}$Cl/Cl.
From the initial set of parameters compatible with B/C, we select
those giving values of $^{10}$Be/$^9$Be and $^{36}$Cl/Cl in the error
bars of {\sc ace} (1$\sigma$ and 3$\sigma$).
As expected from figure~\ref{fig:be_cl}, there is no homogeneous
model which fulfills the above condition at 1$\sigma$
(the two regions delimited by the  filled circles in the left and right
panels do not overlap).
This can be seen in figure~\ref{fig:be_et_cl} which displays the
models compatible with {\sc ace} at the 1$\sigma$ (filled circles)
and 3$\sigma$ (empty circles) level.
We saw in the previous section that high values of $L$ are
independently favored by the two ratios $^{10}$Be/$^9$Be and
$^{36}$Cl/Cl for homogeneous models, and it is natural to recover this
trend in the combined analysis, which points towards $L \gtrsim 6
\unit{kpc}$ at the 3$\sigma$ level.
The allowed range for the other parameters are not much changed.

As regards the influence of $r_{hole}$, we see that the combined
analysis naturally favors hole radii $60 \unit{pc}\lesssim r_{hole}
\lesssim
80 \unit{pc}$ (as seen before, values $r_{hole} \gtrsim 100 \unit{pc}$
are independently excluded at the 3$\sigma$ level), which is in full
agreement with {\sc lism} observations.
Note that for these particular models,
values $L \gtrsim 12 \unit{kpc}$ are excluded at the 1$\sigma$ level.

\subsection{Combined analysis all three radioactive: problem with data
?}

\begin{figure}[hbt!]
\centerline{\includegraphics*[width=0.6\textwidth]{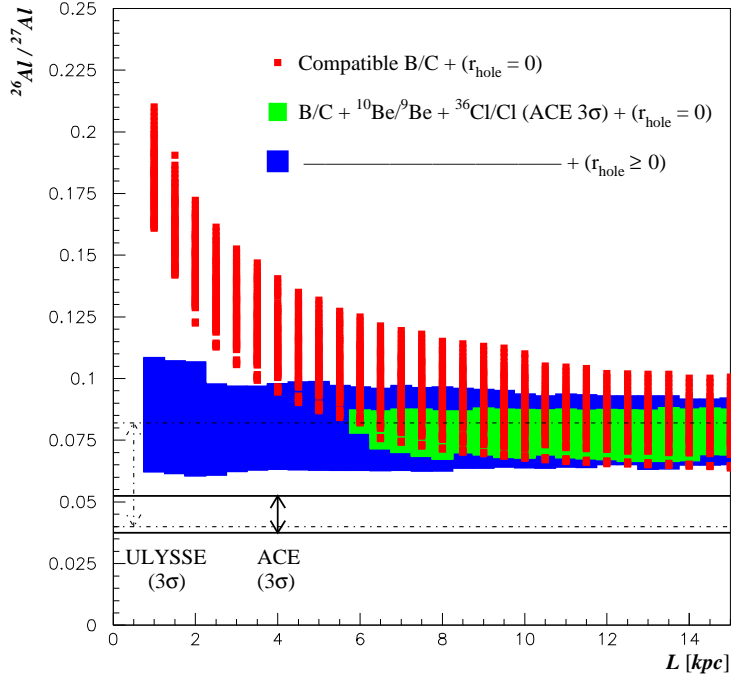}}
\caption{Representation of the $^{26}$Al/$^{27}$Al ratio at the {\sc
ace} energy as a function of $L$ for the models compatible with
(i) B/C in homogeneous models (small squares),
(ii) B/C plus  $^{10}$Be/$^9$Be and $^{36}$Cl/Cl {\sc ace}
3$\sigma$ (medium squares are for homogeneous models and big squares
are for inhomogeneous models).
Solid lines represent the 3$\sigma$ error band from {\sc ace},
dashed lines represent the 3$\sigma$ error band from Ulysses.}
\label{fig:al}
\end{figure}

The next logical step is to include the $^{26}$Al/$^{27}$Al ratio in
the analysis. We first display in figure~\ref{fig:al} the values of this
ratio
for the models discussed in the previous section, {\em i.e.} compatible
with $^{10}$Be/$^9$Be  and $^{36}$Cl/Cl {\sc ace} data at 3$\sigma$.
We also show the $^{26}$Al/$^{27}$Al experimental bounds from {\sc
ace} and Ulysses at 3$\sigma$.

The first strong conclusion is that all these models (homogeneous and
inhomogeneous) give similar
$^{26}$Al/$^{27}$Al values at $\sim 200 \unit{MeV/nuc}$.
Moreover, it is seen that  these values are not compatible
with {\sc ace} data, even at the 6$\sigma$ level.
We will come back to this point later on.
As pointed out in section~\ref{sec:dataset}, other experiments are not
compatible with the {\sc ace} data for this particular ratio.
To consider the possibility that the $^{26}$Al/$^{27}$Al value may be
higher than hinted by the {\sc ace} measurement;
we repeat  the previous analysis applying Ulysses data for this ratio.
\begin{figure}[hbt!]
\centerline{\includegraphics*[width=0.8\textwidth]{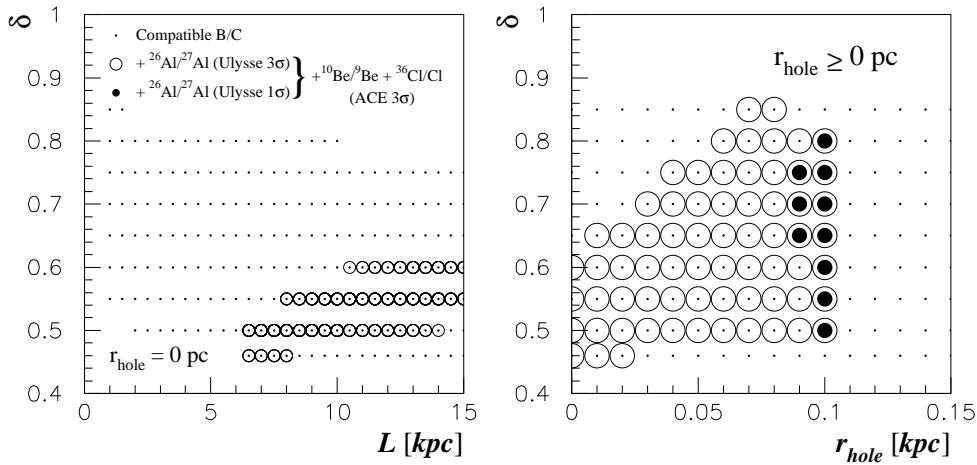}}
\caption{Representation of the models compatible with B/C plus
$^{10}$Be/$^9$Be
and $^{36}$Cl/Cl {\sc ace} 3$\sigma$ plus $^{26}$Al/$^{27}$Al
Ulysses 3$\sigma$ (open circles) and 1$\sigma$ (filled circles) data.
Left panel displays homogeneous models ($r_{hole}=0$) in the plane
$L$--$\delta$.
Right panel displays inhomogeneous models ($r_{hole}\geq 0$) in the
plane
$r_{hole}$--$\delta$.}
\label{fig:the_three_a_la_fois}
\end{figure}

In figure~\ref{fig:the_three_a_la_fois},
we show the models compatible with the $^{10}$Be/$^9$Be and
$^{36}$Cl/Cl {\sc ace} 3$\sigma$ data, plus the $^{26}$Al/$^{27}$Al
Ulysses 3$\sigma$ data (open circles) and 1$\sigma$ data (filled
circles).
In the latter case, only inhomogeneous models with $r_{hole} \approx 100
\unit{pc}$ are consistent with the experimental values for the three
ratios.
In particular, we emphasize that homogeneous models are excluded.

\begin{figure}[hbt!]
\centerline{\includegraphics*[width=0.8\textwidth]{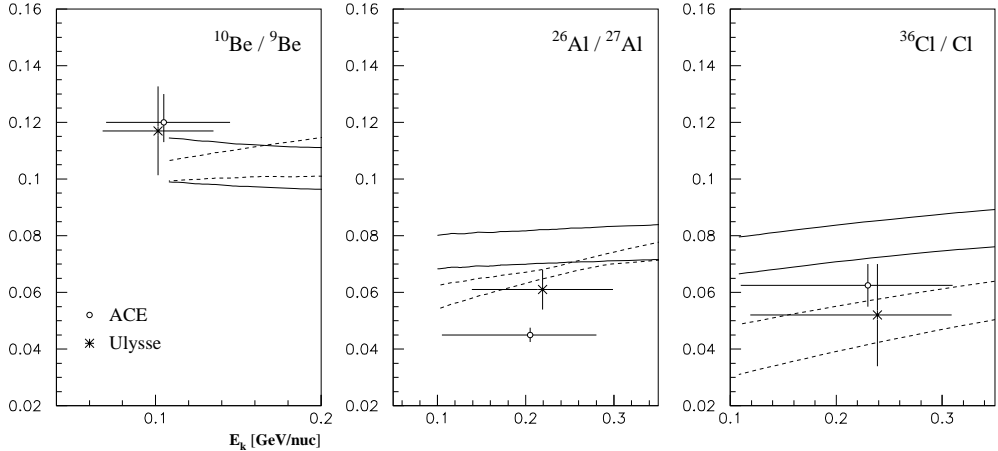}}

\caption{Envelopes of the spectra obtained with all the models
compatible with $^{10}$Be/$^9$Be {\sc ace} and $^{36}$Cl/Cl
{\sc ace} 3$\sigma$ plus $^{26}$Al/$^{27}$Al Ulysses 3$\sigma$
for the three radioactive species.
Solid lines are for homogeneous models ($r_{hole}=0$) and dashed lines
are for inhomogeneous models ($r_{hole} \geq 0$).
Data in the lower panel are from {\sc ace} (circles), Ulysses
(crosses).}
\label{fig:envelopes_be_al_et_cl}
\end{figure}

However, it must be stressed that for these "good" models,
the $^{10}$Be/$^9$Be and $^{36}$Cl/Cl ratios have the lowest possible
value
still compatible with 3$\sigma$ {\sc ace} data, whereas the
$^{26}$Al/$^{27}$Al ratio has the maximal value compatible with
1$\sigma$
Ulysses data, so that these ratio are only marginally consistent.
In figure~\ref{fig:envelopes_be_al_et_cl}, we also present the
envelopes
obtained as in the previous section,
but with the more stringent condition that the $^{10}$Be/$^9$Be and
$^{36}$Cl/Cl ratios are within the 1 $\sigma$ {\sc ace} error bars.

We did not make the analysis with the Voyager data point instead of
Ulysses because the error bars are much larger. However, we notice that
the central value provided by this experiment is in better agreement
with our derived $^{26}$Al/$^{27}$Al ratios (see
figure~\ref{fig:envelopes}).


\section{Conclusion and discussion}

We have presented the analysis of the three radioactive ratios
$^{10}$Be/$^9$Be, $^{36}$Cl/Cl and $^{26}$Al/$^{27}$Al in the
framework of a diffusion model, taking into account the presence of a
local underdensity of radius $r_{hole}$.
We find that the $^{10}$Be/$^9$Be and $^{36}$Cl/Cl ratios are
compatible with {\sc ace} data if $r_{hole} \sim 60-80$ pc, but
that no model is simultaneously compatible with the {\sc ace} data
for the three ratios.
However, if we consider other $^{26}$Al/$^{27}$Al data  (Ulysses or
Voyager), some models are marginally consistent and the cases $60
\unit{pc}\lesssim r_{hole} \lesssim 100 \unit{pc}$ are preferred.
These values are in agreement with independent estimations of {\sc
lism} studies.
The presence of this hole would also lead to an attenuation of the
expected
$^{14}$C flux of the order of $10^{-4}$. Thus, it seems a very
difficult task to detect any $^{14}$C in cosmic radiation, unless some
local source is present.
For example, radioactive nuclei could be produced in the very local
fluff (with a very low density, see section~\ref{sec:lism}).
This has been more realistically modelled by Ptuskin and Soutoul (1990)
and Ptuskin et al. (1997) with a three-layer bubble.
It would be straightforward to adapt our model to
take into account a multi-shell contribution (as long as the symmetry
is preserved).
Moreover, all these models assume that the shells have the same
center, which is
surely wrong, but the generic behaviour in
$\exp(-r_{hole}/l_{rad})$ is expected to hold even in a more
complicated geometry.

We have not yet discussed solar modulation.
All the data have slightly different solar modulation parameters.
We used the simple force field scheme to modulate our fluxes (as in
Paper I and Paper II).
We checked that even taking extreme values for
the modulation parameters yields about $5\%$ changes in calculated
ratio.
This does not affect the results of our analysis.

Our results point out an important feature of cosmic rays. Either there
is a problem with the Al data, or there is a fundamental problem with
the diffusive approach of radioactive nuclei propagation.
Alternatively, there could be a problem with cross sections.
Other similar studies (see for example Strong and Moskalenko, 1998 and
further developments) should be able to confirm our results when
incorporating a hole in their model.


\section*{Acknowledgments}
F.D. gratefully acknowledges a fellowship by the Istituto Nazionale di
Fisica Nucleare. We also would like to thank the French Programme National de
Cosmologie for its financial support.


\appendix

\section{demonstration of formula (\ref{proba_inverse})}

In this section, we compute the probability $d{\cal P}$ that a
radioactive nucleus detected at the center of
the hole has been created by a spallation at a distance between $r$
and $r+dr$, in the framework of the simple modelling of
subsection~\ref{App:hole_phys}.
It is a conditional probability, and as such it can be
expressed, using Bayes theorem, by
\begin{eqnarray}
      d{\cal P}\left\{ \mbox{created at r} | \mbox{detected in 0} \right\}
      &=& {\cal P}\left\{\mbox{detected in 0} | \mbox{created at r} \right\}
      \times \frac{d{\cal P}\left\{ \mbox{created at r} \right\}}
      {{\cal P}\left\{\mbox{detected in 0}\right\}} \nonumber \\
      &\propto& {\cal P}\left\{\mbox{detected in 0} | \mbox{created at
      r} \right\}
      \times d{\cal P}\left\{ \mbox{created at r} \right\}\nonumber
\end{eqnarray}
The first term is given by the expression (\ref{propagateur})
\begin{displaymath}
       {\cal P}\left\{\mbox{detected in 0} | \mbox{created at r} \right\} \propto
       \int_0^\infty \exp \left\{ - \frac{r^2}{4Kt} - \frac{t}{\gamma
       \tau_0} \right\} \frac{dt}{t^{3/2}}
\end{displaymath}
and the second term is simply given by
\begin{displaymath}
     d{\cal P}\left\{ \mbox{created at r} \right\} = 
     \frac{\Theta(r-r_{hole}) \, 2\pi r \, N(r)
     \, dr }{\int 2\pi r N(r) \, dr} \propto \Theta(r-r_{hole}) r N(r)\, dr
\end{displaymath}
where $N(r)$ is the density of parent nuclei. This gives
\begin{displaymath}
    d{\cal P}\left\{ \mbox{created at r} | \mbox{detected in 0}
    \right\} \propto
    \Theta(r-r_{hole}) \, r \, N(r) \, dr \times
    \int_0^\infty \exp \left\{ - \frac{r^2}{4Kt} - \frac{t}{\gamma
    \tau_0} \right\} \frac{dt}{t^{3/2}}
\end{displaymath}
Defining $x=r/\sqrt{4Kt}$, the integral is proportional to
\begin{displaymath}
    \int_0^\infty \exp \left\{ - \frac{r^2}{4K\gamma \tau_0 x^2}
    - x^2 \right\} dx
\end{displaymath}
The value of this integral is given by (\ref{integrale_a_la_con})
which gives
\begin{equation}
       d{\cal P}\left\{ \mbox{created at r} | \mbox{detected in 0} \right\} \propto
       \Theta(r-r_{hole}) \, N(r)\, e^{-r/l_{rad}} dr
       \label{proba_inverse2}
\end{equation}
The proportionality coefficient of this relation would easily be
obtained by imposing that $\int_{r=0}^\infty d{\cal P} = 1$.



\end{document}